\documentclass[prx,aps,twocolumn,superscriptaddress]{revtex4-2}

\usepackage[pdftex]{graphicx}
\usepackage{amsbsy,amssymb,amsmath,bm,mathtools}

\usepackage{xspace}
\usepackage{bm}
\usepackage{color}
 
\usepackage{url}
\usepackage[section]{placeins}
\usepackage[colorlinks=true,linkcolor=blue,citecolor=blue]{hyperref}

\begin{document}


\title{Magnetic HIP-NN for spin dynamics in disordered itinerant magnets}

\author{Supriyo Ghosh}
\affiliation{Department of Physics, University of Virginia, Charlottesville, Virginia 22904, USA}
\affiliation{Department of Chemistry, University of Chicago, Chicago, IL 60637, USA}

\author{Yunhao Fan}
\affiliation{Department of Physics, University of Virginia, Charlottesville, Virginia 22904, USA}

\author{Sheng Zhang}
\affiliation{Department of Physics, University of Virginia, Charlottesville, Virginia 22904, USA}

\author{Kipton Barros}
\affiliation{Theoretical Division and CNLS, Los Alamos National Laboratory, Los Alamos, New Mexico 87545, USA}

\author{Gia-Wei Chern}
\affiliation{Department of Physics, University of Virginia, Charlottesville, Virginia 22904, USA}

\date{\today}

\begin{abstract}
We present a magnetic extension of the Hierarchically Interacting Particle Neural Network (HIP-NN) that enables large-scale simulations of electron-mediated spin dynamics in disordered itinerant magnets. The resulting magnetic HIP-NN (mHIP-NN) incorporates rotationally invariant spin correlations directly into hierarchical message-passing layers, enabling the network to learn emergent magnetic energy landscapes and effective local fields from coupled geometric-spin environments while preserving spin-rotation symmetry. As a benchmark application, we consider structurally disordered itinerant $s$-$d$ exchange models in which the effective magnetic forces arise dynamically from the instantaneous electronic structure and are computationally prohibitive to evaluate using conventional exact-diagonalization-based approaches. We show that mHIP-NN accurately reproduces the local torques governing Landau-Lifshitz-Gilbert dynamics and faithfully captures the nonequilibrium evolution of spatial spin correlations following thermal quenches. Our results establish symmetry-aware hierarchical message-passing networks as an efficient and scalable framework for large-scale simulations of frustrated itinerant spin systems and nonequilibrium magnetic dynamics. More broadly, because the learned energy functional remains fully differentiable with respect to both atomic coordinates and spin variables, the framework also provides a natural foundation for spin-dependent interatomic potentials and coupled atom-spin dynamics.
\end{abstract}

\maketitle

\section{Introduction} \label{sec:intro}

Recent advances in machine-learning (ML) force-field models have transformed large-scale atomistic simulations in chemistry and materials science~\cite{behler07,bartok10,li15,shapeev16,botu17,smith17,zhang18,Lubbers2018,deringer19,drautz19,chmiela17,chmiela18,sauceda20}. By exploiting the locality of electronic structure and learning effective local energy landscapes directly from quantum-mechanical datasets, modern ML frameworks enable molecular and spin-lattice dynamics simulations with near \textit{ab initio} accuracy while retaining computational costs that scale linearly with system size. Early approaches based on handcrafted local descriptors have gradually evolved into graph neural network (GNN) and message-passing architectures that iteratively communicate information between neighboring atoms, allowing the model to automatically learn complex many-body correlations~\cite{scarselli2009,gilmer2017,hamilton2017,xu2019,maron2019,xie2018,schutt2018,choudhary2021,dai2021,reiser2022,batatia2022}. In parallel, equivariant neural networks (ENNs) have emerged as a powerful framework for incorporating geometric and symmetry constraints directly into the network architecture~\cite{cohen2016,cohen2018,weiler2018,kondor2025,batzner2022,kaba2022,musaelian2023,gong2023,batatia2025,yang2025}. Together, these developments have established symmetry-aware message-passing networks as a central paradigm for modern ML interatomic potentials and large-scale molecular dynamics simulations.

Among these developments, the Hierarchically Interacting Particle Neural Network (HIP-NN) introduced an efficient and physically transparent framework for learning local energy decompositions in atomistic systems~\cite{Lubbers2018,Nebgen2018,Smith2019,Zhou2022,Chigaev2023,Zhang2025}. Inspired by many-body expansions, HIP-NN employs alternating interaction and on-site layers to construct hierarchical representations of local atomic environments while naturally preserving translational, rotational, and permutational symmetries. From a modern perspective, the interaction layers of HIP-NN closely resemble the message-passing operations now widely used in graph neural networks. The original HIP-NN framework demonstrated state-of-the-art performance for molecular energy prediction and molecular dynamics benchmarks, while its hierarchical energy decomposition also provided a natural estimate of model uncertainty, since large higher-order corrections indicate stronger dependence on nonlocal many-body correlations that may be underrepresented in the training data. Owing to its physically motivated locality structure and computational efficiency, HIP-NN provides an attractive foundation for scalable ML force fields and extensions to more complex interacting systems.

While machine-learning approaches for magnetic materials have attracted growing attention in recent years~\cite{Eckhoff2021,Kotykhov2024,Yu2024,Yu2024b,Huang2025,Xu2025}, most existing frameworks are primarily designed to construct spin-dependent interatomic potentials or effective magnetic Hamiltonians for equilibrium and finite-temperature materials modeling. In these approaches, the ML models are typically trained to predict energies, magnetic exchange couplings, anisotropies, or related static magnetic properties for prescribed spin configurations or magnetic orders. The resulting models are then mainly employed in molecular dynamics, spin-lattice simulations, or Monte Carlo sampling with the magnetic structure effectively specified or constrained in advance. As such, these methods generally do not aim to learn the nonequilibrium spin dynamics itself, nor the time-dependent effective torques governing dynamical evolution.

A qualitatively different class of ML force-field approaches has recently emerged for nonequilibrium spin dynamics in itinerant magnetic systems, where the effective magnetic forces are generated dynamically by the underlying electronic subsystem itself. In metallic magnets described by itinerant-electron or $s$-$d$-type models, the effective fields entering the Landau-Lifshitz-Gilbert (LLG) equation depend not only on the instantaneous spin configuration, but also on the nonequilibrium response of conduction electrons mediating magnetic interactions. As a result, conventional simulations require repeated electronic-structure calculations throughout the dynamical evolution, rendering large-scale simulations computationally prohibitive. Recent works have shown that these emergent electron-mediated torques can instead be learned directly from microscopic simulations using symmetry-aware descriptors and neural-network force fields~\cite{zhang21,zhang23,cheng23b,Fan24,tyberg25,Chern2026,Chern2026b}, enabling scalable spin-dynamics simulations of correlated lattice models while retaining the essential many-body electronic effects. Extending such ML force-field frameworks to disordered itinerant magnets with off-lattice structures and spatially random interactions, however, remains a significant challenge.

Disordered itinerant magnets provide a particularly challenging setting for electron-mediated spin dynamics. In metallic spin glasses and related systems, quenched disorder and electron-mediated interactions combine to produce complex magnetic energy landscapes with strong frustration and slow relaxation dynamics~\cite{mezard87,binder86,mydosh95,mydosh15,cannella72,lamelas95}. In the weak-coupling limit, effective Ruderman-Kittel-Kasuya-Yosida (RKKY) interactions provide an approximate description of the magnetic forces~\cite{ruderman54,kasuya56,yosida57}. However, in strongly coupled itinerant systems, higher-order electronic processes and disorder-induced localization effects can generate highly nontrivial magnetic interactions that are difficult to capture using simple analytical models or conventional phenomenological spin Hamiltonians. At the same time, the locality of the emergent electronic forces remains sufficiently robust to motivate the construction of scalable local ML force-field models.

In this work, we introduce a magnetic extension of the HIP-NN framework for large-scale simulations of electron-mediated spin dynamics in disordered itinerant magnets. Our approach incorporates rotationally invariant spin correlations directly into the hierarchical interaction layers of HIP-NN, enabling the network to learn emergent electron-mediated magnetic forces from the instantaneous spin environment. To benchmark the method, we consider an amorphous generalization of the itinerant $s$-$d$ model~\cite{Zener1951a,deGennes1960,anderson61} in which localized Heisenberg moments are randomly distributed within a metallic host. Training datasets are generated from exact-diagonalization-based adiabatic spin dynamics simulations, and the resulting magnetic HIP-NN model is shown to accurately reproduce the effective local magnetic fields governing the LLG dynamics while achieving linear computational scaling with system size. We further demonstrate that the locality underlying the ML framework emerges naturally from the spatial decay of the effective electronic interactions in the disordered itinerant system.

\section{Magnetic Extension of the HIP-NN Framework}
\label{sec:m-hip-nn}

\begin{figure*}
\centering
\includegraphics[width=1.99\columnwidth]{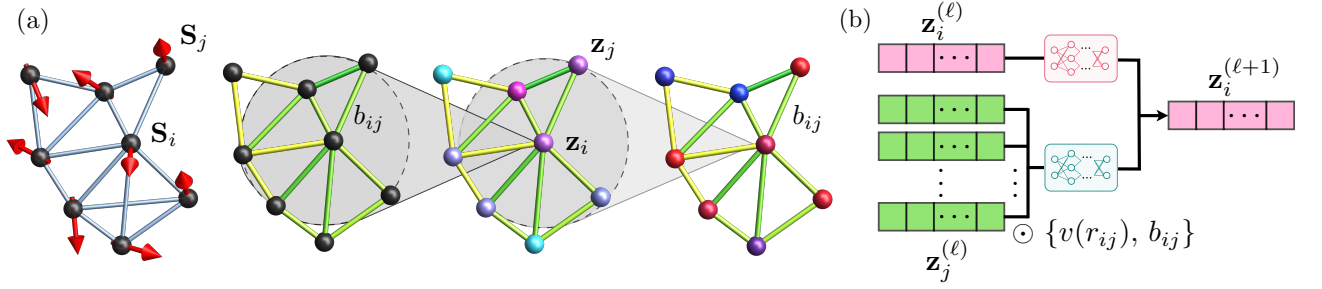}
\caption{
Schematic illustration of the magnetic HIP-NN framework for electron-mediated spin dynamics in disordered itinerant magnets. 
(a) Localized Heisenberg spins $\{\mathbf S_i\}$ embedded in an amorphous lattice are mapped onto rotationally invariant bond features $b_{ij} = \mathbf S_i \cdot \mathbf S_j$, which are combined with geometric information associated with inter-site distances $r_{ij}$. These quantities define the local environments used to construct hierarchical feature representations $\{\mathbf z_i\}$ on each magnetic site. 
(b) Within the interaction layer, neighboring feature vectors $\mathbf z_j^{(\ell)}$ are weighted by learned distance-dependent functions $v(r_{ij})$ together with the spin-correlation features $b_{ij}$, and are subsequently processed through nonlinear neural-network mappings to generate updated site features $\mathbf z_i^{(\ell+1)}$. Repeated interaction and on-site layers enable the network to learn effective electron-mediated magnetic interactions and the local fields governing LLG spin dynamics.
}
\label{fig:HIP-NN-schematic}
\end{figure*}

The magnetic HIP-NN (mHIP-NN) framework developed in this work is designed for magnetic materials in which both atomic and spin degrees of freedom contribute to the emergent energy landscape. In such systems, the total energy depends simultaneously on the geometric configuration of atoms and the orientations of localized magnetic moments. Consequently, the underlying symmetry structure differs fundamentally from that of conventional atomistic ML force fields. In addition to the Euclidean symmetry group $E(3)$ associated with translations, rotations, and reflections in real space, magnetic systems also possess an internal spin-rotation symmetry $SO(3)$ associated with global rotations of the magnetic moments. In many systems, particularly in the weak spin-orbit-coupling limit, these geometric and spin sectors are approximately decoupled. The ML architecture must therefore incorporate both spatial and spin symmetries in a manner consistent with the microscopic Hamiltonian.

The mHIP-NN framework is designed to provide a symmetry-aware representation of magnetic materials and disordered spin systems characterized by the configuration $\mathcal C = \left\{ \mathbf r_i,\, Z_i,\, \mathbf S_i \right\}$, where $\mathbf r_i$ denotes the atomic position of site $i$, $Z_i$ labels the local magnetic species or internal spin type, and $\mathbf S_i$ specifies the orientation of the corresponding magnetic moment. The species label $Z_i$ may encode microscopic attributes such as atomic identity, spin magnitude, or local magnetic class, while the spin vectors $\mathbf S_i$ describe the orientational degrees of freedom associated with the magnetic subsystem. In contrast to conventional atomistic ML models, where the interaction layers depend only on geometric information associated with the local atomic environment, the present magnetic extension additionally incorporates rotationally invariant spin correlations between neighboring magnetic moments. The resulting architecture therefore generalizes hierarchical message passing from purely geometric environments to coupled geometric-spin systems relevant for magnetic materials and spin-lattice dynamics.

The geometric structure of the system is represented through pairwise distances 
\begin{equation}
	r_{ij} = |\mathbf r_i-\mathbf r_j|,
\end{equation}
which are invariant under global translations, rotations, and reflections in real space, thereby ensuring consistency with the underlying $E(3)$ symmetry of the atomic environment. Following the locality principle underlying modern ML force fields, only neighboring pairs satisfying $r_{ij} \le R_c$ are retained, where $R_c$ is a finite cutoff radius. In practice, the interaction kernels are further multiplied by smooth cutoff functions to ensure continuity and differentiability of the predicted energies and generalized forces with respect to atomic coordinates.

To incorporate the independent global $SO(3)$ spin-rotation symmetry, the interaction layers must additionally be constructed from rotationally invariant combinations of spins. In the present work, the fundamental magnetic descriptors are chosen to be the bond variables~\cite{zhang21,zhang23}
\begin{equation}
	b_{ij} = \mathbf S_i \cdot \mathbf S_j,
\label{eq:bond_variable}
\end{equation}
which characterize the local spin correlations between neighboring moments. Importantly, these scalar bond variables remain invariant under simultaneous global rotations of all spins and therefore provide a natural symmetry-preserving representation of the magnetic environment. Physically, the quantities $b_{ij}$ encode the relative alignment between neighboring moments, allowing the ML model to distinguish locally ferromagnetic, antiferromagnetic, and noncollinear spin textures while remaining independent of the global spin reference frame.

A schematic of the mHIP-NN is shown in Fig.~\ref{fig:HIP-NN-schematic}. From the perspective of modern GNNs, the magnetic system may be viewed as a graph in which magnetic sites correspond to nodes and neighboring pairs within the cutoff radius $R_c$ define the edges. The architecture of mHIP-NN consists of multiple hidden layers acting on the same underlying graph structure, while progressively constructing increasingly refined representations of the local magnetic environment through iterative message-passing operations. At each layer $\ell$, every site $i$ is associated with a feature vector 
\begin{eqnarray}
	\mathbf z_i^{(\ell)} = \bigl(z^{(\ell)}_{i,1}, \, z^{(\ell)}_{i,2}, \cdots, z^{(\ell)}_{i,N_f} \bigr) \in \mathbb R^{N_f}, 
\end{eqnarray}
where $N_f$ is the number of latent features.

The input feature vector $\mathbf z_i^{(0)}$ encodes the local species information $Z_i$, analogous to node embeddings in modern GNN architectures. As information propagates through successive hidden layers, the node features $\mathbf z_i^{(\ell)}$ become progressively dressed by the surrounding geometric and magnetic environments, thereby allowing the network to learn increasingly complex many-body correlations associated with electron-mediated magnetic interactions. In this sense, the latent variables $\mathbf z_i^{(\ell)}$ should not be interpreted as handcrafted descriptors, but rather as emergent learned representations generated self-consistently through hierarchical message passing on the coupled geometric-spin graph.

\subsection{Propagation rules and message passing}

The hidden layers of mHIP-NN iteratively update the node features $\mathbf z_i^{(\ell)}$ through a sequence of nonlinear transformations defined on the underlying geometric-spin graph. The propagation rules consist of two complementary operations: local on-site transformations acting independently on each node, and interaction layers that communicate information between neighboring sites through message passing. Repeated application of these operations progressively constructs increasingly refined representations of the local magnetic environment and its surrounding geometric-spin correlations.

We first consider the on-site transformations, which act independently on each node without communication between different sites. The on-site update is given by
\begin{eqnarray}
	\tilde z_{i,\alpha}^{(\ell+1)} = \sigma\left( \sum_\beta W_{\alpha\beta}^{(\ell)} z_{i,\beta}^{(\ell)} + c_\alpha^{(\ell)} \right),
\end{eqnarray}
where $\alpha,\beta$ denote feature indices, $W_{\alpha\beta}^{(\ell)}$ is a trainable weight matrix acting in the latent feature space, and $c_\alpha^{(\ell)}$ is a trainable bias parameter. These on-site layers perform local nonlinear transformations of the latent node features while preserving the locality structure of the graph.

Several activation functions may in principle be employed within the hidden layers. In the present work, we adopt the softplus activation function~\cite{Lubbers2018}
\begin{eqnarray}
	\sigma(x) = \ln(1+e^x),
\end{eqnarray}
following the original HIP-NN formulation. Compared to piecewise-linear activations such as ReLU, the softplus function provides a smooth nonlinear mapping, which is advantageous for constructing differentiable energy landscapes and the associated generalized forces derived from the ML energy functional.

The transformed features are subsequently combined with the original features through the residual update
\begin{eqnarray}
	z_{i,\alpha}^{(\ell+1)} = \sum_\beta \tilde W_{\alpha\beta}^{(\ell)} \tilde z_{i,\beta}^{(\ell+1)}
	+ z_{i,\alpha}^{(\ell)} + \tilde c_\alpha^{(\ell)},
\end{eqnarray}
where $\tilde W_{\alpha\beta}^{(\ell)}$ and $\tilde c_\alpha^{(\ell)}$ denote an additional trainable linear transformation and bias parameter. This residual structure is closely related to the ResNet architecture widely used in deep neural networks~\cite{He2016}, in which skip connections allow the transformed features to be added back to the original input features. Rather than reconstructing the entire latent representation at each layer, the network learns only incremental corrections to the features generated at the previous stage. Such residual connections substantially improve the trainability and stability of deep architectures by facilitating the propagation of information and gradients across many hidden layers.

The interaction layers perform the central message-passing operation of mHIP-NN by transmitting information between neighboring sites. At each interaction layer, neighboring feature vectors are first aggregated into local messages according to
\begin{equation}
	m_{i,\alpha}^{(\ell)}
	=
	\sum_{j \in \mathcal N(i)}
	\sum_\beta
	v_{\alpha\beta}^{(\ell)}(r_{ij}, b_{ij})
	z_{j,\beta}^{(\ell)},
\label{eq:message_aggregation}
\end{equation}
where $\mathcal N(i)$ denotes the set of neighboring sites satisfying $r_{ij}\le R_c$. The interaction kernel $v_{\alpha\beta}^{(\ell)}(r_{ij},b_{ij})$ depends explicitly on both the pairwise distances $r_{ij}$ and the spin-invariant bond variables $b_{ij}=\mathbf S_i\cdot\mathbf S_j$, thereby allowing the propagated messages to encode both geometric and local magnetic correlations.

To parameterize the interaction kernels, we expand them in terms of a set of trainable radial sensitivity functions,
\begin{equation}
	v_{\alpha\beta}^{(\ell)}(r_{ij},b_{ij})
	=
	\sum_{\nu=1}^{N_s}
	V_{\nu,\alpha\beta}^{(\ell)} s_\nu(r_{ij})
	+
	\sum_{\nu=1}^{N_s}
	\tilde V_{\nu,\alpha\beta}^{(\ell)}
	s_\nu(r_{ij})\, b_{ij},
\label{eq:v_expansion}
\end{equation}
where $N_s$ denotes the number of sensitivity functions. The first term describes purely geometric message passing, while the second term introduces a spin-dependent modulation through the bond variables $b_{ij}$, such that the effective interaction strength transmitted across each edge depends jointly on the geometric separation and the local spin alignment of neighboring sites. The radial sensitivity functions are chosen as
\begin{equation}
	s_\nu(r)
	=
	\exp\left[
	-
	\frac{(r^{-1}-\mu_\nu^{-1})^2}
	{2\sigma_\nu^{-2}}
	\right]
	\phi_{\rm cut}(r),
\label{eq:sensitivity}
\end{equation}
where $\mu_\nu$ and $\sigma_\nu$ are trainable parameters controlling the center and width of the radial filters. The cutoff function is given by $\phi_{\rm cut}(r)=\cos^2\left(\frac{\pi r}{2R_c}\right)$ for $r\le R_c$ and zero otherwise.

The aggregated messages are subsequently combined with the local node features through the nonlinear update
\begin{equation}
	z_{i,\alpha}^{(\ell+1)}
	=
	\sigma\left[
	m_{i,\alpha}^{(\ell)}
	+
	\sum_\beta
	W_{\alpha\beta}^{(\ell)}
	z_{i,\beta}^{(\ell)}
	+
	c_\alpha^{(\ell)}
	\right],
\label{eq:interaction_layer}
\end{equation}
thereby defining the updated latent representation at the next hidden layer. Repeated application of the interaction and on-site layers progressively enlarges the effective receptive field of the node features $\mathbf z_i^{(\ell)}$. Consequently, deeper hidden layers encode increasingly nonlocal many-body geometric-spin correlations generated through multiple message-passing steps across the graph.

An important property of this interaction-layer construction is that it naturally preserves the global time-reversal symmetry commonly present in magnetic systems. Since the message-passing kernel depends on the bond variables $b_{ij}=\mathbf S_i\cdot\mathbf S_j$, which are bilinear in the spin variables, the interaction kernels remain invariant under the global spin inversion $\mathbf S_i\rightarrow -\mathbf S_i$. Consequently, the propagated messages and the resulting latent representations automatically respect the time-reversal symmetry of the underlying magnetic Hamiltonian, without requiring additional symmetry-enforcement procedures or data augmentation.

\subsection{Hierarchical energy decomposition and force-field formulation}

The primary objective of mHIP-NN is to construct an accurate differentiable representation of the total energy landscape associated with the coupled geometric-spin configuration of the system. Following the original HIP-NN formulation, the total energy is expressed as a sum of local energy contributions associated with individual magnetic sites,
\begin{equation}
	E = \sum_i E_i .
\end{equation}
This locality-based decomposition provides a natural framework for constructing transferable and size-extensive ML force fields, while simultaneously allowing the network to learn increasingly nonlocal many-body correlations through the hierarchical message-passing architecture.

The local energies are constructed from the learned node features $z_i^{(\ell)}$ generated throughout the hidden layers of the network. In contrast to many conventional graph neural network architectures, where observables are typically predicted solely from the final hidden layer, HIP-NN employs a hierarchical energy decomposition in which multiple hidden layers contribute directly to the energy prediction. Specifically, at selected hidden layers $\ell_n$, the node features are linearly projected onto partial energy contributions
\begin{equation}
	E_i^{(n)} = \sum_\alpha w_\alpha^{(n)} z_{i,\alpha}^{(\ell_n)} + c^{(n)},
\end{equation}
where $\alpha$ labels the feature index, while $w_\alpha^{(n)}$ and $c^{(n)}$ are trainable parameters. The total local energy is then obtained from the sum of hierarchical contributions
\begin{equation}
	E_i = \sum_{n=0}^{N_{\rm interaction}} E_i^{(n)} .
\end{equation}

The hierarchical index $n$ reflects the progressively increasing complexity of the effective local environment encoded by successive hidden layers of the network. From the perspective of message passing on the geometric-spin graph, repeated interaction layers systematically enlarge the receptive field of the node features $z_{i,\alpha}^{(\ell)}$, allowing deeper layers to incorporate increasingly nonlocal many-body correlations. Consequently, lower-order contributions primarily capture local geometric-spin correlations associated with nearby neighbors, while higher-order terms encode more complex emergent correlations generated through multiple message-passing steps across the graph. Importantly, all hierarchical contributions are generated coherently within a single ML architecture, enabling the network to self-consistently learn effective representations of electron-mediated magnetic interactions directly from the underlying geometric and spin environments. This hierarchical decomposition is one of the characteristic features of the HIP-NN framework and differs from more conventional GNN architectures in which observables are constructed solely from the final hidden layer representation.

An important advantage of this formulation is that the total energy $E$ remains fully differentiable with respect to both the atomic coordinates $\{\mathbf r_i\}$ and spin variables $\{\mathbf S_i\}$. As a result, the corresponding generalized forces can be obtained directly through automatic differentiation of the ML energy functional.

Differentiation with respect to the atomic coordinates yields the atomic forces
\begin{equation}
	\mathbf F_i = - \frac{\partial E}{\partial \mathbf r_i},
\end{equation}
which govern the lattice dynamics of the system. This aspect closely parallels the original HIP-NN framework developed for ML interatomic potentials and force-field models, where the neural network is trained to construct a differentiable representation of the underlying potential-energy surface. Because the total energy depends on the geometric environment through the message-passing architecture and radial sensitivity functions, the resulting forces naturally incorporate complex many-body geometric correlations learned from the training data. These forces may subsequently be employed within conventional molecular dynamics or Langevin simulations of atomic motion.

In the present magnetic extension of HIP-NN, however, the total energy additionally depends explicitly on the local spin variables through the spin-dressed interaction kernels and bond variables $b_{ij} = \mathbf S_i \cdot \mathbf S_j$. Consequently, differentiation with respect to the spin degrees of freedom produces effective local magnetic fields
\begin{equation}
	\mathbf H_i  = - \frac{\partial E}{\partial \mathbf S_i},
\end{equation}
which provide the driving forces for spin dynamics. Physically, these effective fields encode the complex electron-mediated magnetic interactions learned implicitly by the network from the underlying geometric-spin configurations. In contrast to conventional spin Hamiltonians with explicitly prescribed exchange couplings, the present ML formulation allows the effective magnetic interactions to emerge directly from the hierarchical nonlinear representation learned by the network.

The effective magnetic fields $\mathbf H_i$ naturally enter LLG-type equations governing the nonequilibrium dynamics of the local magnetic moments. The present framework therefore provides a unified ML formulation for coupled atom-spin dynamics, in which both lattice and magnetic degrees of freedom evolve self-consistently on the learned energy landscape. In the present work, however, we focus specifically on spin dynamics in metallic spin-glass systems, where the atomic positions are fixed and the dominant dynamical variables are the local magnetic moments.

\section{Application to disordered itinerant magnets}

\label{sec:results}

In this section, we apply the mHIP-NN framework to the nonequilibrium dynamics of disordered metallic spin systems. Our primary objective is to investigate whether the ML model can accurately learn the emergent effective energy landscape and electronic driving forces generated by itinerant electrons in structurally disordered magnetic environments relevant to metallic spin glasses. To this end, we first introduce a generic class of disordered \textit{s-d} exchange models describing itinerant-electron-mediated spin interactions, followed by a specific disordered tight-binding realization used as the benchmark system in the present work.

\subsection{Disordered \textit{s-d} exchange models}

The starting point is a generic \textit{s-d} exchange model describing localized magnetic moments coupled to conduction electrons,
\begin{align}
   \hat{{\mathcal{H}}}=  \hat{{\mathcal{H}}}_e(\hat{c}, \hat{c}^\dagger) -J\sum_i \hat{\bold{S}}_i \cdot \hat{\bm{s}}_i,
\label{H1}
\end{align} 
where
\begin{align}
	\hat{\bm{s}}_i = \frac{\hbar}{2} (\hat{c}^\dagger_{i,\alpha}\bm{\sigma}_{\alpha \beta} \hat{c}_{i,\beta})
\end{align}
denotes the spin density of itinerant conduction electrons, and $\hat{c}_{i,\alpha}^\dagger$ ($\hat{c}_{i,\alpha}$) creates (annihilates) an electron with spin $\alpha=\uparrow,\downarrow$ at site $i$. The vectors $\hat{\bold{S}}_i$ represent localized classical magnetic moments, while the coupling constant $J$ describes the local exchange interaction between itinerant electrons and impurity spins. Depending on the form of the electronic Hamiltonian $\hat{\mathcal H}_e$, Eq.~(\ref{H1}) encompasses a broad class of metallic magnetic systems, including Kondo-lattice models, double-exchange systems, and disordered metallic spin glasses.

In the weak-coupling limit, the conduction electrons may be perturbatively integrated out, leading to effective oscillatory interactions between local moments of the RKKY type. In disordered systems, the random spatial arrangement of magnetic impurities naturally generates frustrated interactions of varying signs and strengths, producing the complex energy landscape characteristic of metallic spin glasses. In realistic systems, however, especially in the intermediate- or strong-coupling regime, the effective magnetic interactions generally become substantially more complicated than simple pairwise RKKY couplings. Higher-order electronic processes and nonlinear multi-spin interactions can no longer be neglected, and the resulting effective spin Hamiltonian acquires a highly nontrivial dependence on both the geometric disorder and the instantaneous spin configuration. One of the central motivations of the present work is therefore to investigate whether the mHIP-NN framework can efficiently learn these emergent effective interactions directly from microscopic electronic calculations.

Our primary interest is the nonequilibrium dynamics of the localized magnetic moments, which is governed by the stochastic LLG equation~\cite{Landau1935,Gilbert2004,Evans2014}
\begin{align}
    \frac{d\bold{S}_i}{dt} = \gamma \bold{S}_i \times (\bm{\bold{H}}_i+\bm{\eta}_i)-\alpha \bold{S}_i\times (\bold{S}_i\times \bold{H}_i),
\label{eq:LLG}
\end{align}
where $\gamma$ is the gyromagnetic ratio, $\alpha$ is the damping coefficient, $\bm{\eta}_i$ denotes stochastic thermal noise, and $\bold{H}_i$ is the effective local magnetic field acting on spin $\bold{S}_i$. Within the adiabatic approximation, the electronic subsystem relaxes on a timescale much faster than the spin dynamics, such that the electrons remain approximately in a quasistatic equilibrium state corresponding to the instantaneous spin configuration. The effective spin energy is therefore obtained by integrating out the electronic degrees of freedom,
\begin{align}
	E = \langle \hat{\mathcal H} \rangle = {\rm Tr}(\hat\rho_e \hat{\mathcal H}),
\end{align}
where $\hat \rho_e$ is the equilibrium electronic density matrix. The effective magnetic field entering the LLG equation is then obtained from the Hellmann-Feynman force~\cite{Feynman1939}
\begin{align}
	\bold{H}_i = - \Bigl\langle \frac{\partial \hat{\mathcal{H}}}{\partial \mathbf S_i} \Bigr\rangle.
\end{align}
In conventional approaches, evaluating these forces generally requires repeated large-scale electronic-structure calculations throughout the dynamical evolution, which rapidly becomes computationally prohibitive for large systems or long simulation times. The present systems therefore provide an ideal benchmark for the mHIP-NN framework, where the ML model aims to learn the effective electronic forces directly from microscopic training data.

To provide a concrete realization of the above generic disordered \textit{s-d} spin-glass model, we consider in this work a structurally disordered tight-binding system with distance-dependent electron hopping. Unlike conventional Kondo-lattice models, where local moments occupy periodic lattice sites, the localized spins $\bold{S}_i$ are distributed randomly throughout a three-dimensional cubic simulation box of side length $L$. To avoid unphysical clustering, a minimum-distance constraint is imposed such that $r_{ij}=|\bm r_i-\bm r_j|>r_{\rm min}$ for all pairs of spins.

For a given realization of the random atomic configuration $\{\mathbf r_i\}$, the electronic subsystem is described by a disordered tight-binding Hamiltonian with a hopping amplitude decaying exponentially with the distance between spins
\begin{eqnarray}
	\hat{\mathcal{H}}_e = \sum_{ij} \sum_{\alpha=\uparrow\downarrow} t_0 \,e^{-|\mathbf r_i - \mathbf r_j| / l}\, \hat c_{i,\alpha}^\dagger \hat c^{\,}_{j,\alpha} 
	\label{H2}
\end{eqnarray}
where $t_0$ defines the characteristic electronic energy scale and $l$ controls the spatial decay length of the hopping process. Tight-binding models with exponentially decaying hopping amplitudes of this form have been widely employed in the study of amorphous and structurally disordered electronic systems. The random spatial arrangement of sites naturally generates strong disorder in the electronic subsystem, which in turn produces highly nontrivial effective magnetic interactions between localized moments.

The effective magnetic field entering the LLG dynamics is obtained from the instantaneous electronic state through the Hellmann-Feynman relation
\begin{align}
	\bold{H}_i = J\hbar/2 \sum_{\alpha \beta} \bm{\sigma}_{\alpha \beta} C_{i\alpha,j\beta}.
\end{align}
where $C_{i\alpha,j\beta}$ is the single-electron density matrix or correlation function,
\begin{align}
	C_{i\alpha,j\beta} = \langle \hat{c}^\dagger_{i,\alpha} \hat{c}_{j,\beta} \rangle
\end{align}
which, in principle, must be evaluated repeatedly throughout the spin dynamics. In conventional approaches, this typically requires repeated exact diagonalization or other large-scale electronic calculations at every time step. The present disordered tight-binding realization of the generic \textit{s-d} spin-glass model therefore provides a particularly stringent test of the mHIP-NN framework, since the ML model must learn the complex effective energy landscape generated by the interplay between geometric disorder, electron hopping, and spin-electron coupling.

\begin{figure}
\centering
\includegraphics[width=0.99\columnwidth]{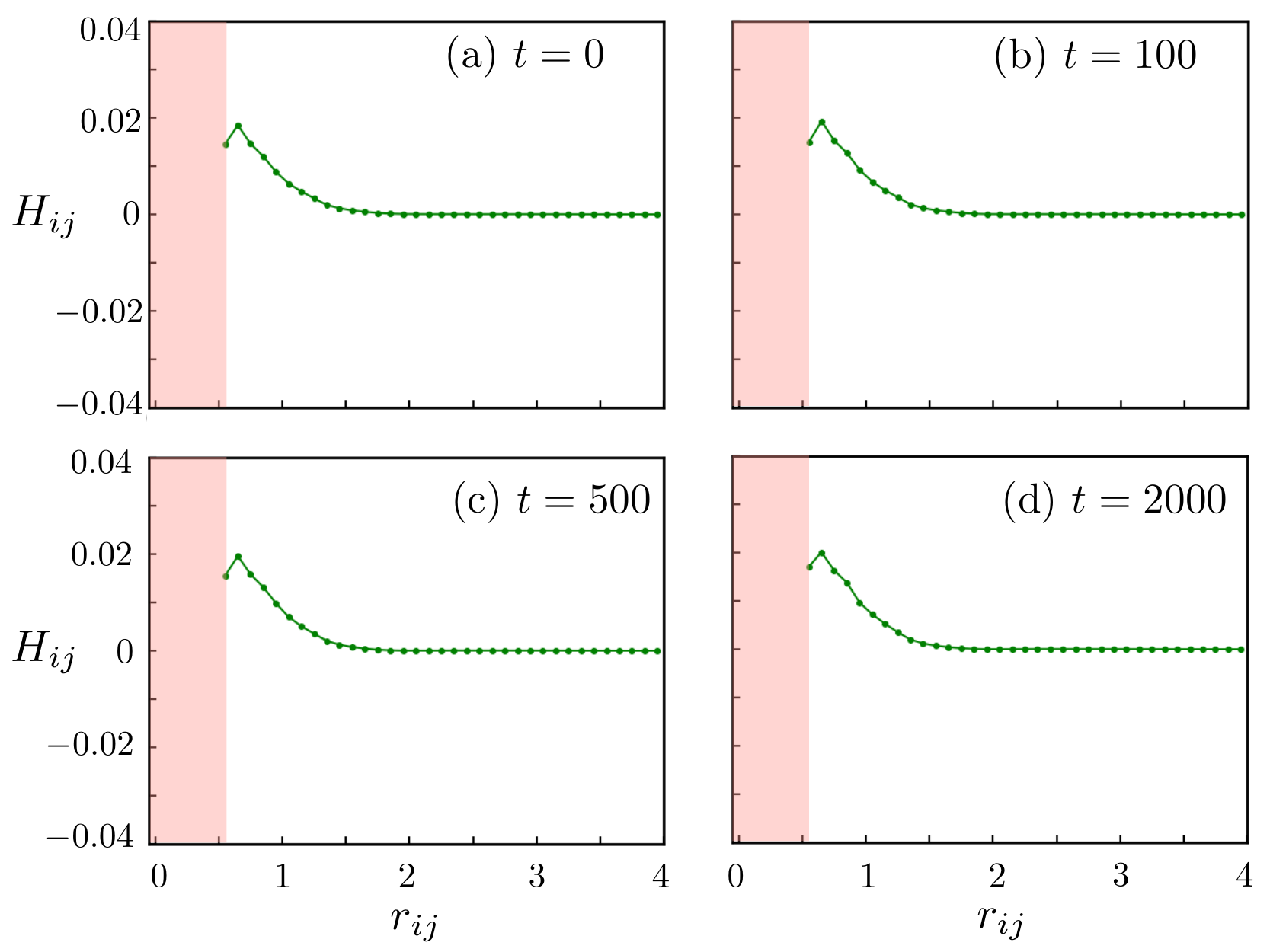}
\caption{Spatial dependence of the Hessian matrix elements $H_{ij}=\partial^2 E/\partial S_i^x \partial S_j^x$ evaluated from the ML-generated energy functional at different times during the ED-LLG simulations: (a) $t=0$, (b) $t=100$, (c) $t=500$, and (d) $t=2000$. The Hessian elements are plotted as a function of the intersite distance $r_{ij}$. The shaded region denotes the forbidden short-distance regime $r_{ij}<r_{\rm min}$ in the underlying spin-glass model. The rapid decay of $H_{ij}$ demonstrates the locality and stability of the learned magnetic interactions throughout the dynamical evolution.  }
\label{fig:hessian}
\end{figure}

The locality of the effective spin Hamiltonian is a key ingredient underlying the efficiency and scalability of the present ML framework. To quantify the spatial range of the effective magnetic interactions, we examine the Hessian matrix $H_{ij}=d^2E/dS_i dS_j$, where the derivatives are taken with respect to the spin orientations under the fixed-length constraint $|\mathbf S_i|=1$. As shown in Fig.~\ref{fig:hessian}, the Hessian elements obtained from ED-LLG simulations for the representative parameters $J=6t_0$ and simulation box size $L=5l$ decay rapidly with increasing intersite distance $r_{ij}$ and become negligibly small beyond a characteristic range of approximately $1.5l$. This behavior indicates that, although the effective spin-spin interaction originates from itinerant electrons and is formally long-ranged, it remains sufficiently local for the construction of accurate linear-scaling ML force-field models. The effective locality is further enhanced by the intrinsic structural randomness of the underlying tight-binding spin-glass model, which can induce partial localization of the electronic wave functions through Anderson-localization effects, thereby suppressing long-distance magnetic couplings.

\subsection{Construction and Training of the mHIP-NN Model}

The mHIP-NN model is trained using datasets generated from exact-diagonalization-based adiabatic spin-dynamics simulations of the disordered itinerant spin models introduced in the previous subsection. For each spin configuration $\{\mathbf S_i\}$, the electronic Hamiltonian is diagonalized numerically to obtain the total energy and the corresponding effective magnetic fields acting on the local moments, which serve as the reference labels for supervised learning.

In the present implementation, the network consists of $N_{\rm int}=2$ interaction layers, each followed by $N_{\rm os}=3$ on-site layers with residual connections. A fixed feature dimension $N_f=80$ is used throughout the hidden layers after the initial nonlinear embedding of the scalar inputs into higher-dimensional feature channels. The interaction layers encode the environmental dependence arising from neighboring spins and spatial geometry, while the on-site layers further process the local features through nonlinear transformations. Similar to the original HIP-NN architecture, the hierarchical organization of the hidden layers enables the network to systematically capture correlations across multiple length scales. In addition, the interaction layers incorporate the spin-dependent coupling factors $v_{ab}^{(\ell)}$, ensuring that the network preserves the global SO(3) spin-rotation symmetry.

All trainable parameters are initialized using Xavier uniform initialization, including the weights $w$, $W$, $\tilde W$, $V$, and $\tilde V$, while all bias parameters are initialized to zero. Throughout the hidden layers, we employ the smooth nonlinear activation function softplus, following the original HIP-NN implementation. To stabilize the hierarchical energy decomposition during training, the hierarchical energy weights are additionally rescaled using the energy variance of the training dataset. 

For the radial sensitivity functions appearing in the interaction layers, we use the cutoff radius $R_c=2.5l$ and  $N_{\rm sensitivity}=20$ sensitivity channels parameterized by $\mu_{\nu,l}$ and $\sigma_{\nu,l}$ ($\nu=1,\cdots,N_{\rm sensitivity}$). Their initial values are chosen using the hyperparameters $R_{\rm low}=0.28$ and $R_{\rm high}=1.67$, such that the inverse centers $\mu_{\nu,l}^{-1}$ are uniformly distributed between $R_{\rm high}^{-1}$ and $R_{\rm low}^{-1}$, while the widths are initialized to a fixed value proportional to $N_{\rm sensitivity}R_{\rm low}$.

\begin{figure}
\centering
\includegraphics[width=0.99\columnwidth]{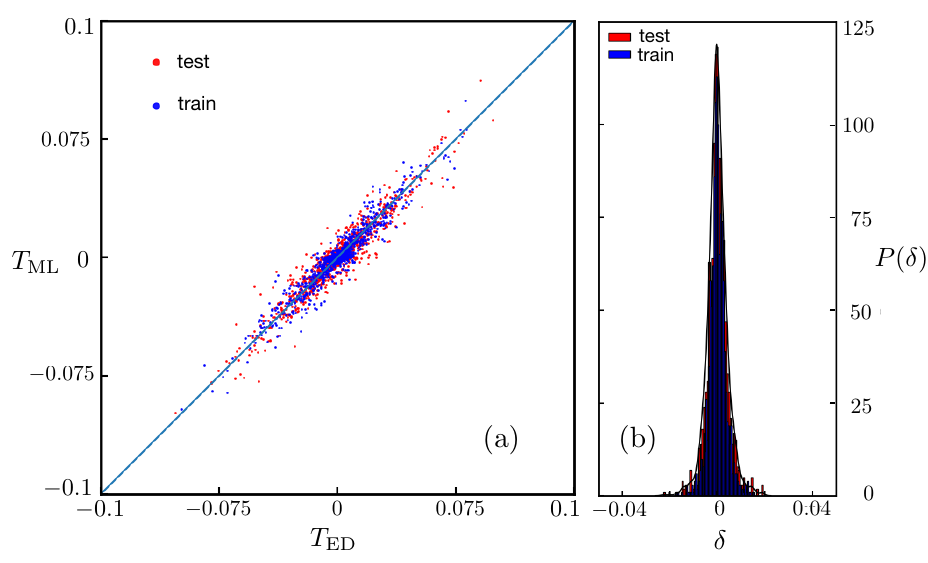}
\caption{(a) Parity plot of the mHIP-NN predicted torque components,  $\mathbf{T}^{\rm ML}_i=\mathbf{S}_i\times\mathbf{H}^{\rm ML}_i$, versus the exact ED results for the training (blue) and test (red) datasets.  Each point corresponds to one Cartesian component of the local torque vector.  (b) Distribution of the prediction error  $\delta = T^{\rm ML}-T^{\rm ED}$ for the training and test sets, showing a narrow distribution centered around zero.  }
\label{fig:benchmark1}
\end{figure}

The effective magnetic fields entering the LLG dynamics are obtained through automatic differentiation of the neural-network energy functional with respect to the local spin variables using the PyTorch framework. Rather than directly minimizing the errors of the local fields themselves, we construct the loss function using the local torques $\mathbf T_i = \mathbf S_i \times \mathbf H_i$, since only the transverse component of the effective field contributes to spin precession and relaxation dynamics. In particular, during dissipative relaxation the spins tend to remain nearly aligned with their local effective fields, making the perpendicular component the more dynamically relevant quantity. The loss function is therefore defined as
\begin{equation}
	\mathcal L = \sum_{i=1}^{N} \left| \mathbf T_{i}^{\rm ED} - 	\mathbf T_{i}^{\rm ML} \right|^2
	+ \eta_E \left| E^{\rm ED} - E^{\rm ML} \right|^2 ,
\end{equation}
where the parameter $\eta_E$ controls the relative weight between the torque and energy contributions during training.

The model parameters are optimized using the Adam stochastic optimizer~\cite{kingma17} with a learning rate of $10^{-5}$. To improve generalization and reduce overfitting, we further employ early stopping and five-fold cross-validation during the training procedure.

We now discuss the generation of the training dataset used for the mHIP-NN model. Here we focus on the disordered $s$-$d$ Hamiltonian in Eq.~(\ref{H2}) with a strong electron--spin coupling $J=6t_0$, well beyond the perturbative regime where the RKKY approximation is expected to hold. In this strong-coupling regime, the effective spin interactions become highly nonlinear and intrinsically many-body in nature, making analytical approaches intractable and motivating the development of a ML force-field framework for large-scale spin dynamics. To generate the training data, we performed ED-LLG simulations for systems containing $N=100$ randomly distributed magnetic atoms with half-filled electron bands inside a cubic box of linear size $L=5l$. For numerical stability, a minimum interatomic distance of $0.5l$ was imposed between all pairs of atoms. The dataset was constructed from $10$ independent random atomic configurations, each combined with $10$ independent initial spin configurations. During the relaxation dynamics of each simulation, $500$ spin configurations were sampled and used as training snapshots. Since the local effective field $\mathbf H_i$ acting on every spin serves as an independent training target, the total effective dataset size becomes $500\times10\times10\times100 = 5\times10^6$ local training samples.

\subsection{Benchmarks of mHIP-NN}

Figure~\ref{fig:benchmark1} demonstrates the strong predictive performance of the mHIP\mbox{-}NN model for the local torque $\mathbf{T}_i=\mathbf{S}_i\times\mathbf{H}_{i}$ throughout the spin dynamics. As shown in the parity plot of Fig.~\ref{fig:benchmark1}(a), the predicted torque components closely follow the exact ED results for both the training and test datasets, with only small deviations from the diagonal line. The corresponding error distributions in Fig.~\ref{fig:benchmark1}(b) are sharply peaked around zero and exhibit very small variance, indicating both high accuracy and excellent generalization capability of the model. From the test dataset, we estimate a coefficient of determination of approximately $R^2 \sim 0.98$ and a mean-squared error of order ${\rm MSE}\sim 10^{-5}$. The close agreement demonstrates that the mHIP\mbox{-}NN architecture, formulated in terms of the bond variables, successfully captures the essential local magnetic interactions and their associated torques.

\begin{figure}
\centering
\includegraphics[width=0.99\columnwidth]{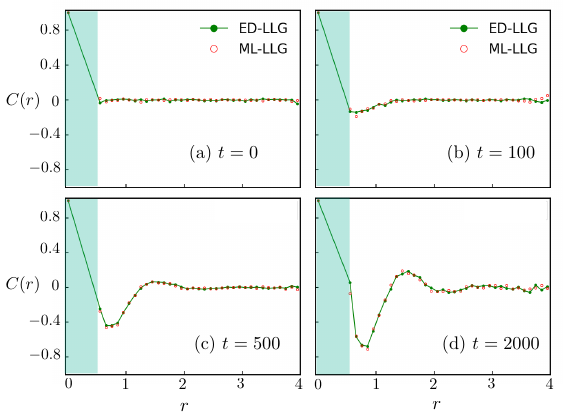}
\caption{Spin-spin correlation function $C(r)$ at different times following a thermal quench, obtained from ED-LLG and ML-LLG simulations: (a) $t=0$, (b) $t=100$, (c) $t=500$, and (d) $t=2000$. The close agreement between the ED and ML results demonstrates that the learned force-field model accurately reproduces the nonequilibrium spin dynamics and the evolution of spatial magnetic correlations. }
\label{fig:corr}
\end{figure}

Having established the accuracy of the mHIP\mbox{-}NN model in predicting the instantaneous local torques, we now proceed to a more stringent dynamical benchmark by directly integrating the trained ML force field into the LLG simulations. At each timestep, the effective local magnetic fields are generated from the ML-predicted energy functional and subsequently used to evolve the spin dynamics according to Eq.~(\ref{eq:LLG}). The central question is whether the learned force field can faithfully reproduce not only the instantaneous driving forces, but also the emergent nonequilibrium evolution of the interacting spin system over long timescales.

To benchmark the dynamical performance, we perform thermal-quench simulations of the disordered $s$-$d$ spin model from random initial spin configurations to a low temperature $T=0.001$. Both the ED-LLG and ML-LLG simulations employ the same damping coefficient $\alpha=0.05\gamma$, where $\gamma$ is the gyromagnetic ratio. For each realization, the simulations are initialized using the same disordered atomic configuration and random spin state, and the results are further averaged over multiple disorder and spin realizations. To characterize the evolving magnetic textures, we compute the equal-time spin-spin correlation function
\begin{equation}
	C(r_{ij},t) = \langle \mathbf S_i(t)\cdot\mathbf S_j(t)\rangle - \langle \mathbf S_i(t)\rangle \cdot 	\langle \mathbf S_j(t)\rangle,
\end{equation}
where $\langle\cdots\rangle$ denotes the ensemble average over both atomic disorder and random initial spin configurations.

The resulting correlation functions are shown in Fig.~\ref{fig:corr}. Throughout the entire relaxation process, the ML-LLG results exhibit nearly perfect agreement with the corresponding ED-LLG simulations, demonstrating that the learned force field accurately reproduces the nonequilibrium spin dynamics and the evolution of spatial magnetic correlations. During the early stages of relaxation, a pronounced short-range antiferromagnetic correlation develops near the minimum allowed separation $r_{\rm min}=0.5l$, reflecting the dominant local antiferromagnetic coupling mediated by the itinerant electrons. As the system further relaxes toward lower-energy glassy states, weaker longer-range oscillatory correlations gradually emerge at intermediate distances. At late times, the correlation functions approach a nearly stationary form with vanishing net magnetization, indicating the relaxation toward a frozen disordered spin-glass state. The excellent agreement between the ED and ML simulations across the entire time evolution therefore demonstrates that the mHIP\mbox{-}NN force field successfully captures both the local magnetic interactions and the emergent collective dynamics of the itinerant spin system.

\begin{figure}
\centering
\includegraphics[width=0.99\columnwidth]{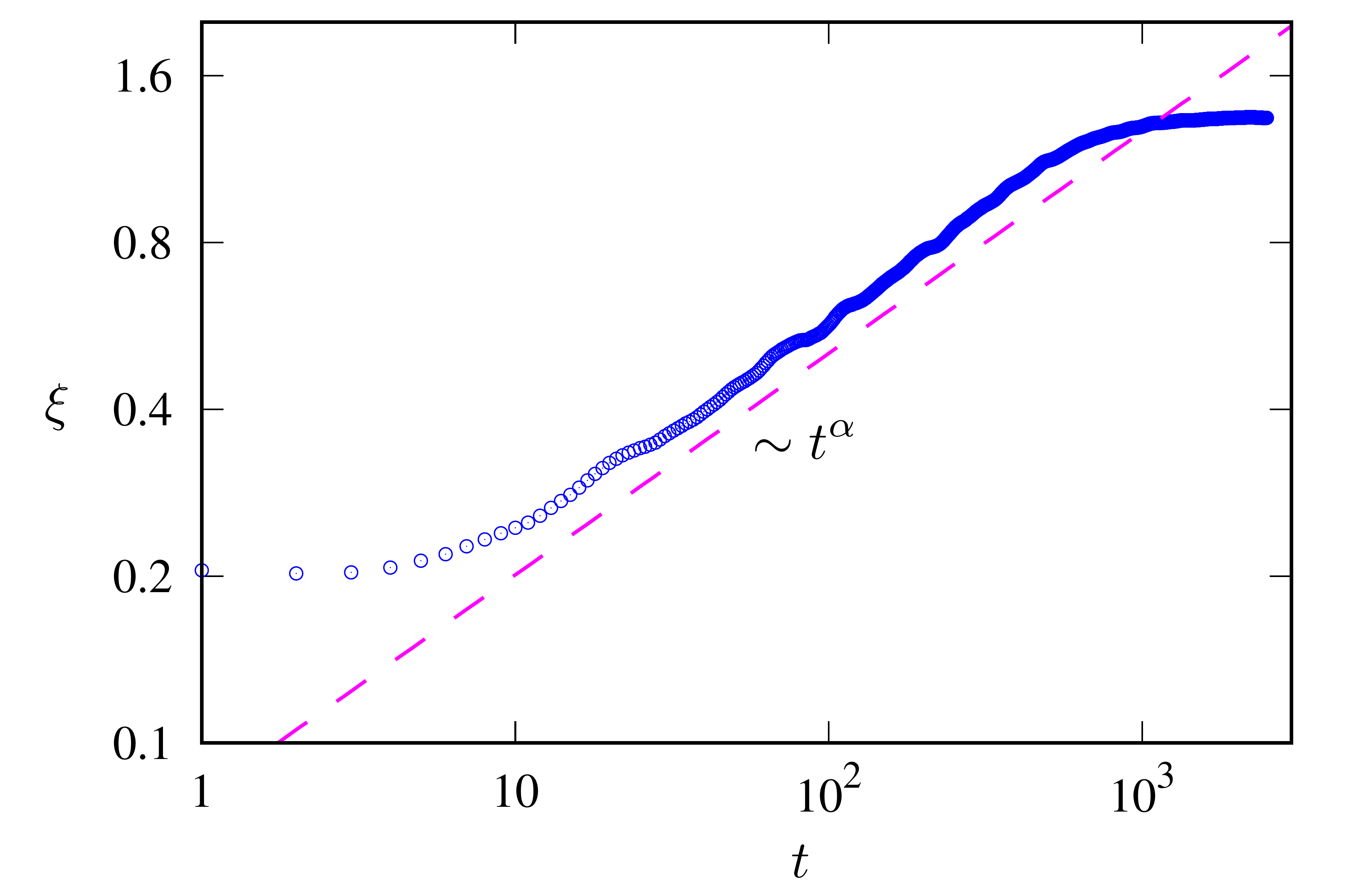}
\caption{
Time evolution of the spin-glass correlation length $\xi(t)$ following a thermal quench in the disordered itinerant spin model. The correlation length exhibits a slow coarsening dynamics characterized by an approximate power-law growth, $\xi(t) \sim t^{\alpha}$, over an extended intermediate-time regime, as indicated by the dashed guide line with exponent $\alpha \approx 0.4$. At longer times, the growth gradually slows down and begins to saturate due to finite-size effects.
}
\label{fig:xi}
\end{figure}

To further quantify the emergent spatial correlations during the nonequilibrium relaxation process, we extract a characteristic correlation length $\xi(t)$ from the equal-time spin-spin correlation function through the first moment of the absolute correlation profile, $\xi(t) = \int_0^\infty r |C(r,t)|dr \big/ \int_0^\infty |C(r,t)|dr$.
Fig.~\ref{fig:xi} shows the time evolution of $\xi(t)$ following the thermal quench. After a short initial transient, the correlation length exhibits a slow algebraic increase over an extended intermediate-time regime,
\begin{equation}
	\xi(t) \sim t^{\alpha},
\end{equation}
with an effective growth exponent $\alpha \approx 0.4$. At longer times, the growth gradually slows down and eventually begins to saturate due to finite-size effects.

The observed increase of $\xi(t)$ indicates the progressive emergence of spatially correlated spin textures during the relaxation process. Unlike conventional phase-ordering kinetics in ferromagnets or antiferromagnets, however, the present disordered itinerant spin system does not develop a simple symmetry-broken long-range ordered state. Consequently, the growing correlation length should not be interpreted as the size of conventional magnetic domains separated by domain walls. Instead, $\xi(t)$ characterizes the spatial extent over which the spins become dynamically correlated through the slow relaxation dynamics within the frustrated energy landscape generated by the disorder and the oscillatory electron-mediated magnetic interactions.

In this sense, the relaxation dynamics exhibits a coarsening-like evolution of correlated spin structures, despite the absence of conventional long-range magnetic order. Similar growing dynamical length scales extracted from equal-time spin-spin correlations have been widely employed in studies of aging and slow relaxation in disordered and frustrated spin systems. Here, we do not attempt a full characterization of an equilibrium spin-glass phase. Nevertheless, the observed slow growth of the correlation length, together with the absence of long-range magnetic order, is consistent with the emergence of slow glassy relaxation dynamics in the frustrated itinerant spin system.

The emergence of the power-law growth of $\xi(t)$ further highlights the highly collective nature of the nonequilibrium dynamics in the disordered itinerant spin model. Since the correlation length is constructed from the evolving many-body spin textures over a broad range of length and time scales, the ability of the ML-driven dynamics to reproduce the underlying spatial correlations and relaxation behaviors shown in Fig.~\ref{fig:corr} demonstrates that the mHIP\mbox{-}NN force field successfully captures the cooperative dynamical evolution of the frustrated spin system beyond merely reproducing instantaneous local torques.

\section{Conclusion and Outlook}

In summary, we have developed a magnetic extension of the HIP-NN framework for large-scale simulations of electron-mediated spin dynamics in disordered itinerant magnets. By incorporating rotationally invariant spin correlations directly into hierarchical message-passing layers, the resulting mHIP-NN architecture provides a symmetry-aware representation of coupled geometric-spin environments and enables the construction of differentiable magnetic energy functionals from microscopic electronic data. Using structurally disordered itinerant $s$-$d$ exchange models as benchmark systems, we demonstrated that the learned ML force field accurately reproduces the local torques and nonequilibrium Landau-Lifshitz-Gilbert dynamics generated by exact electronic calculations. In particular, the mHIP-NN framework successfully captures the evolution of spatial spin correlations and the slow coarsening-like growth of dynamical correlation lengths associated with glassy relaxation dynamics in frustrated itinerant spin systems.

The present work opens a promising route toward realistic large-scale simulations of metallic spin glasses and disordered magnetic materials with emergent electron-mediated interactions. Owing to the computational complexity of repeatedly solving the underlying electronic problem, many previous studies of spin-glass dynamics have relied on phenomenological lattice spin models with simplified short-range random couplings or Monte Carlo-based surrogate dynamics. By contrast, the present framework provides a systematic route for constructing scalable off-lattice ML force fields directly from microscopic itinerant-electron Hamiltonians. Since the resulting ML model effectively replaces repeated electronic-structure calculations with a computationally efficient differentiable energy functional, substantially larger spatial and temporal scales become accessible for studies of nonequilibrium relaxation, aging, coarsening, and driven dynamics in frustrated metallic magnets. More broadly, the framework suggests a possible route toward realistic simulations of dilute magnetic alloys and structurally disordered magnetic materials, where effective electronic Hamiltonians derived from density-functional-theory or other electronic-structure approaches could be combined with ML force-field training to enable large-scale dynamical simulations beyond the reach of direct electronic calculations.

Beyond spin dynamics in itinerant magnets, the mHIP-NN formulation also provides a natural foundation for spin-dependent interatomic potentials and coupled atom-spin dynamics. Because the learned energy functional remains fully differentiable with respect to both atomic coordinates and spin variables, the same framework can simultaneously generate atomic forces and effective magnetic fields through automatic differentiation. This opens the possibility of extending modern ML interatomic potentials to systems in which lattice and magnetic degrees of freedom evolve self-consistently, including magnetic phase transitions, ultrafast spin-lattice relaxation, magnetoelastic phenomena, and nonequilibrium structural dynamics in magnetic materials. In this broader context, the present work may serve as a step toward unified symmetry-aware ML force fields for large-scale simulations of coupled structural, electronic, and magnetic dynamics in complex materials.

\bigskip

\begin{acknowledgments}
This work was supported by the U.S. Department of Energy, Office of Science, Office of Basic Energy Sciences, under Award No. DE-SC0020330. K.B. acknowledges support from the U.S. Department of Energy, Office of Science, Office of Basic Energy Sciences through Los Alamos National Laboratory. G.-W. Chern thanks Nicholas Lubbers for many insightful discussions. The authors acknowledge Research Computing at the University of Virginia for providing computational resources and technical support.\end{acknowledgments}

\bibliography{ref}

\begin{thebibliography}{70}%
\makeatletter
\providecommand \@ifxundefined [1]{%
 \@ifx{#1\undefined}
}%
\providecommand \@ifnum [1]{%
 \ifnum #1\expandafter \@firstoftwo
 \else \expandafter \@secondoftwo
 \fi
}%
\providecommand \@ifx [1]{%
 \ifx #1\expandafter \@firstoftwo
 \else \expandafter \@secondoftwo
 \fi
}%
\providecommand \natexlab [1]{#1}%
\providecommand \enquote  [1]{``#1''}%
\providecommand \bibnamefont  [1]{#1}%
\providecommand \bibfnamefont [1]{#1}%
\providecommand \citenamefont [1]{#1}%
\providecommand \href@noop [0]{\@secondoftwo}%
\providecommand \href [0]{\begingroup \@sanitize@url \@href}%
\providecommand \@href[1]{\@@startlink{#1}\@@href}%
\providecommand \@@href[1]{\endgroup#1\@@endlink}%
\providecommand \@sanitize@url [0]{\catcode `\\12\catcode `\$12\catcode
  `\&12\catcode `\#12\catcode `\^12\catcode `\_12\catcode `\%12\relax}%
\providecommand \@@startlink[1]{}%
\providecommand \@@endlink[0]{}%
\providecommand \url  [0]{\begingroup\@sanitize@url \@url }%
\providecommand \@url [1]{\endgroup\@href {#1}{\urlprefix }}%
\providecommand \urlprefix  [0]{URL }%
\providecommand \Eprint [0]{\href }%
\providecommand \doibase [0]{https://doi.org/}%
\providecommand \selectlanguage [0]{\@gobble}%
\providecommand \bibinfo  [0]{\@secondoftwo}%
\providecommand \bibfield  [0]{\@secondoftwo}%
\providecommand \translation [1]{[#1]}%
\providecommand \BibitemOpen [0]{}%
\providecommand \bibitemStop [0]{}%
\providecommand \bibitemNoStop [0]{.\EOS\space}%
\providecommand \EOS [0]{\spacefactor3000\relax}%
\providecommand \BibitemShut  [1]{\csname bibitem#1\endcsname}%
\let\auto@bib@innerbib\@empty
\bibitem [{\citenamefont {Behler}\ and\ \citenamefont
  {Parrinello}(2007)}]{behler07}%
  \BibitemOpen
  \bibfield  {author} {\bibinfo {author} {\bibfnamefont {J.}~\bibnamefont
  {Behler}}\ and\ \bibinfo {author} {\bibfnamefont {M.}~\bibnamefont
  {Parrinello}},\ }\bibfield  {title} {\bibinfo {title} {Generalized
  neural-network representation of high-dimensional potential-energy
  surfaces},\ }\href {https://doi.org/10.1103/PhysRevLett.98.146401} {\bibfield
   {journal} {\bibinfo  {journal} {Phys. Rev. Lett.}\ }\textbf {\bibinfo
  {volume} {98}},\ \bibinfo {pages} {146401} (\bibinfo {year}
  {2007})}\BibitemShut {NoStop}%
\bibitem [{\citenamefont {Bart\'ok}\ \emph {et~al.}(2010)\citenamefont
  {Bart\'ok}, \citenamefont {Payne}, \citenamefont {Kondor},\ and\
  \citenamefont {Cs\'anyi}}]{bartok10}%
  \BibitemOpen
  \bibfield  {author} {\bibinfo {author} {\bibfnamefont {A.~P.}\ \bibnamefont
  {Bart\'ok}}, \bibinfo {author} {\bibfnamefont {M.~C.}\ \bibnamefont {Payne}},
  \bibinfo {author} {\bibfnamefont {R.}~\bibnamefont {Kondor}},\ and\ \bibinfo
  {author} {\bibfnamefont {G.}~\bibnamefont {Cs\'anyi}},\ }\bibfield  {title}
  {\bibinfo {title} {Gaussian approximation potentials: The accuracy of quantum
  mechanics, without the electrons},\ }\href
  {https://doi.org/10.1103/PhysRevLett.104.136403} {\bibfield  {journal}
  {\bibinfo  {journal} {Phys. Rev. Lett.}\ }\textbf {\bibinfo {volume} {104}},\
  \bibinfo {pages} {136403} (\bibinfo {year} {2010})}\BibitemShut {NoStop}%
\bibitem [{\citenamefont {Li}\ \emph {et~al.}(2015)\citenamefont {Li},
  \citenamefont {Kermode},\ and\ \citenamefont {De~Vita}}]{li15}%
  \BibitemOpen
  \bibfield  {author} {\bibinfo {author} {\bibfnamefont {Z.}~\bibnamefont
  {Li}}, \bibinfo {author} {\bibfnamefont {J.~R.}\ \bibnamefont {Kermode}},\
  and\ \bibinfo {author} {\bibfnamefont {A.}~\bibnamefont {De~Vita}},\
  }\bibfield  {title} {\bibinfo {title} {Molecular dynamics with on-the-fly
  machine learning of quantum-mechanical forces},\ }\href
  {https://doi.org/10.1103/PhysRevLett.114.096405} {\bibfield  {journal}
  {\bibinfo  {journal} {Phys. Rev. Lett.}\ }\textbf {\bibinfo {volume} {114}},\
  \bibinfo {pages} {096405} (\bibinfo {year} {2015})}\BibitemShut {NoStop}%
\bibitem [{\citenamefont {Shapeev}(2016)}]{shapeev16}%
  \BibitemOpen
  \bibfield  {author} {\bibinfo {author} {\bibfnamefont {A.~V.}\ \bibnamefont
  {Shapeev}},\ }\bibfield  {title} {\bibinfo {title} {Moment tensor potentials:
  A class of systematically improvable interatomic potentials},\ }\href
  {https://doi.org/10.1137/15M1054183} {\bibfield  {journal} {\bibinfo
  {journal} {Multiscale Modeling \& Simulation}\ }\textbf {\bibinfo {volume}
  {14}},\ \bibinfo {pages} {1153} (\bibinfo {year} {2016})}\BibitemShut
  {NoStop}%
\bibitem [{\citenamefont {Botu}\ \emph {et~al.}(2017)\citenamefont {Botu},
  \citenamefont {Batra}, \citenamefont {Chapman},\ and\ \citenamefont
  {Ramprasad}}]{botu17}%
  \BibitemOpen
  \bibfield  {author} {\bibinfo {author} {\bibfnamefont {V.}~\bibnamefont
  {Botu}}, \bibinfo {author} {\bibfnamefont {R.}~\bibnamefont {Batra}},
  \bibinfo {author} {\bibfnamefont {J.}~\bibnamefont {Chapman}},\ and\ \bibinfo
  {author} {\bibfnamefont {R.}~\bibnamefont {Ramprasad}},\ }\bibfield  {title}
  {\bibinfo {title} {Machine learning force fields: Construction, validation,
  and outlook},\ }\href {https://doi.org/10.1021/acs.jpcc.6b10908} {\bibfield
  {journal} {\bibinfo  {journal} {The Journal of Physical Chemistry C}\
  }\textbf {\bibinfo {volume} {121}},\ \bibinfo {pages} {511} (\bibinfo {year}
  {2017})}\BibitemShut {NoStop}%
\bibitem [{\citenamefont {Smith}\ \emph {et~al.}(2017)\citenamefont {Smith},
  \citenamefont {Isayev},\ and\ \citenamefont {Roitberg}}]{smith17}%
  \BibitemOpen
  \bibfield  {author} {\bibinfo {author} {\bibfnamefont {J.~S.}\ \bibnamefont
  {Smith}}, \bibinfo {author} {\bibfnamefont {O.}~\bibnamefont {Isayev}},\ and\
  \bibinfo {author} {\bibfnamefont {A.~E.}\ \bibnamefont {Roitberg}},\
  }\bibfield  {title} {\bibinfo {title} {Ani-1: an extensible neural network
  potential with dft accuracy at force field computational cost},\ }\href
  {https://doi.org/10.1039/C6SC05720A} {\bibfield  {journal} {\bibinfo
  {journal} {Chem. Sci.}\ }\textbf {\bibinfo {volume} {8}},\ \bibinfo {pages}
  {3192} (\bibinfo {year} {2017})}\BibitemShut {NoStop}%
\bibitem [{\citenamefont {Zhang}\ \emph {et~al.}(2018)\citenamefont {Zhang},
  \citenamefont {Han}, \citenamefont {Wang}, \citenamefont {Car},\ and\
  \citenamefont {E}}]{zhang18}%
  \BibitemOpen
  \bibfield  {author} {\bibinfo {author} {\bibfnamefont {L.}~\bibnamefont
  {Zhang}}, \bibinfo {author} {\bibfnamefont {J.}~\bibnamefont {Han}}, \bibinfo
  {author} {\bibfnamefont {H.}~\bibnamefont {Wang}}, \bibinfo {author}
  {\bibfnamefont {R.}~\bibnamefont {Car}},\ and\ \bibinfo {author}
  {\bibfnamefont {W.}~\bibnamefont {E}},\ }\bibfield  {title} {\bibinfo {title}
  {Deep potential molecular dynamics: A scalable model with the accuracy of
  quantum mechanics},\ }\href {https://doi.org/10.1103/PhysRevLett.120.143001}
  {\bibfield  {journal} {\bibinfo  {journal} {Phys. Rev. Lett.}\ }\textbf
  {\bibinfo {volume} {120}},\ \bibinfo {pages} {143001} (\bibinfo {year}
  {2018})}\BibitemShut {NoStop}%
\bibitem [{\citenamefont {Lubbers}\ \emph {et~al.}(2018)\citenamefont
  {Lubbers}, \citenamefont {Smith},\ and\ \citenamefont
  {Barros}}]{Lubbers2018}%
  \BibitemOpen
  \bibfield  {author} {\bibinfo {author} {\bibfnamefont {N.}~\bibnamefont
  {Lubbers}}, \bibinfo {author} {\bibfnamefont {J.~S.}\ \bibnamefont {Smith}},\
  and\ \bibinfo {author} {\bibfnamefont {K.}~\bibnamefont {Barros}},\
  }\bibfield  {title} {\bibinfo {title} {Hierarchical modeling of molecular
  energies using a deep neural network},\ }\href
  {https://doi.org/10.1063/1.5011181} {\bibfield  {journal} {\bibinfo
  {journal} {The Journal of Chemical Physics}\ }\textbf {\bibinfo {volume}
  {148}},\ \bibinfo {pages} {241715} (\bibinfo {year} {2018})}\BibitemShut
  {NoStop}%
\bibitem [{\citenamefont {Deringer}\ \emph {et~al.}(2019)\citenamefont
  {Deringer}, \citenamefont {Caro},\ and\ \citenamefont {Csanyi}}]{deringer19}%
  \BibitemOpen
  \bibfield  {author} {\bibinfo {author} {\bibfnamefont {V.~L.}\ \bibnamefont
  {Deringer}}, \bibinfo {author} {\bibfnamefont {M.~A.}\ \bibnamefont {Caro}},\
  and\ \bibinfo {author} {\bibfnamefont {G.}~\bibnamefont {Csanyi}},\
  }\bibfield  {title} {\bibinfo {title} {Machine learning interatomic
  potentials as emerging tools for materials science},\ }\href
  {https://doi.org/10.1002/adma.201902765} {\bibfield  {journal} {\bibinfo
  {journal} {Advanced Materials}\ }\textbf {\bibinfo {volume} {31}},\ \bibinfo
  {pages} {1902765} (\bibinfo {year} {2019})}\BibitemShut {NoStop}%
\bibitem [{\citenamefont {Drautz}(2019)}]{drautz19}%
  \BibitemOpen
  \bibfield  {author} {\bibinfo {author} {\bibfnamefont {R.}~\bibnamefont
  {Drautz}},\ }\bibfield  {title} {\bibinfo {title} {Atomic cluster expansion
  for accurate and transferable interatomic potentials},\ }\href
  {https://doi.org/10.1103/PhysRevB.99.014104} {\bibfield  {journal} {\bibinfo
  {journal} {Phys. Rev. B}\ }\textbf {\bibinfo {volume} {99}},\ \bibinfo
  {pages} {014104} (\bibinfo {year} {2019})}\BibitemShut {NoStop}%
\bibitem [{\citenamefont {Chmiela}\ \emph {et~al.}(2017)\citenamefont
  {Chmiela}, \citenamefont {Tkatchenko}, \citenamefont {Sauceda}, \citenamefont
  {Poltavsky}, \citenamefont {Schütt},\ and\ \citenamefont
  {Müller}}]{chmiela17}%
  \BibitemOpen
  \bibfield  {author} {\bibinfo {author} {\bibfnamefont {S.}~\bibnamefont
  {Chmiela}}, \bibinfo {author} {\bibfnamefont {A.}~\bibnamefont {Tkatchenko}},
  \bibinfo {author} {\bibfnamefont {H.~E.}\ \bibnamefont {Sauceda}}, \bibinfo
  {author} {\bibfnamefont {I.}~\bibnamefont {Poltavsky}}, \bibinfo {author}
  {\bibfnamefont {K.~T.}\ \bibnamefont {Schütt}},\ and\ \bibinfo {author}
  {\bibfnamefont {K.-R.}\ \bibnamefont {Müller}},\ }\bibfield  {title}
  {\bibinfo {title} {Machine learning of accurate energy-conserving molecular
  force fields},\ }\href {https://doi.org/10.1126/sciadv.1603015} {\bibfield
  {journal} {\bibinfo  {journal} {Science Advances}\ }\textbf {\bibinfo
  {volume} {3}},\ \bibinfo {pages} {e1603015} (\bibinfo {year}
  {2017})}\BibitemShut {NoStop}%
\bibitem [{\citenamefont {Chmiela}\ \emph {et~al.}(2018)\citenamefont
  {Chmiela}, \citenamefont {Sauceda}, \citenamefont {M{\"u}ller},\ and\
  \citenamefont {Tkatchenko}}]{chmiela18}%
  \BibitemOpen
  \bibfield  {author} {\bibinfo {author} {\bibfnamefont {S.}~\bibnamefont
  {Chmiela}}, \bibinfo {author} {\bibfnamefont {H.~E.}\ \bibnamefont
  {Sauceda}}, \bibinfo {author} {\bibfnamefont {K.-R.}\ \bibnamefont
  {M{\"u}ller}},\ and\ \bibinfo {author} {\bibfnamefont {A.}~\bibnamefont
  {Tkatchenko}},\ }\bibfield  {title} {\bibinfo {title} {Towards exact
  molecular dynamics simulations with machine-learned force fields},\ }\href
  {https://doi.org/10.1038/s41467-018-06169-2} {\bibfield  {journal} {\bibinfo
  {journal} {Nature Communications}\ }\textbf {\bibinfo {volume} {9}},\
  \bibinfo {pages} {3887} (\bibinfo {year} {2018})}\BibitemShut {NoStop}%
\bibitem [{\citenamefont {Sauceda}\ \emph {et~al.}(2020)\citenamefont
  {Sauceda}, \citenamefont {Gastegger}, \citenamefont {Chmiela}, \citenamefont
  {Müller},\ and\ \citenamefont {Tkatchenko}}]{sauceda20}%
  \BibitemOpen
  \bibfield  {author} {\bibinfo {author} {\bibfnamefont {H.~E.}\ \bibnamefont
  {Sauceda}}, \bibinfo {author} {\bibfnamefont {M.}~\bibnamefont {Gastegger}},
  \bibinfo {author} {\bibfnamefont {S.}~\bibnamefont {Chmiela}}, \bibinfo
  {author} {\bibfnamefont {K.-R.}\ \bibnamefont {Müller}},\ and\ \bibinfo
  {author} {\bibfnamefont {A.}~\bibnamefont {Tkatchenko}},\ }\bibfield  {title}
  {\bibinfo {title} {{Molecular force fields with gradient-domain machine
  learning (GDML): Comparison and synergies with classical force fields}},\
  }\href {https://doi.org/10.1063/5.0023005} {\bibfield  {journal} {\bibinfo
  {journal} {The Journal of Chemical Physics}\ }\textbf {\bibinfo {volume}
  {153}},\ \bibinfo {pages} {124109} (\bibinfo {year} {2020})}\BibitemShut
  {NoStop}%
\bibitem [{\citenamefont {Scarselli}\ \emph {et~al.}(2009)\citenamefont
  {Scarselli}, \citenamefont {Gori}, \citenamefont {Tsoi}, \citenamefont
  {Hagenbuchner},\ and\ \citenamefont {Monfardini}}]{scarselli2009}%
  \BibitemOpen
  \bibfield  {author} {\bibinfo {author} {\bibfnamefont {F.}~\bibnamefont
  {Scarselli}}, \bibinfo {author} {\bibfnamefont {M.}~\bibnamefont {Gori}},
  \bibinfo {author} {\bibfnamefont {A.~C.}\ \bibnamefont {Tsoi}}, \bibinfo
  {author} {\bibfnamefont {M.}~\bibnamefont {Hagenbuchner}},\ and\ \bibinfo
  {author} {\bibfnamefont {G.}~\bibnamefont {Monfardini}},\ }\bibfield  {title}
  {\bibinfo {title} {The graph neural network model},\ }\href
  {https://doi.org/10.1109/TNN.2008.2005605} {\bibfield  {journal} {\bibinfo
  {journal} {IEEE Transactions on Neural Networks}\ }\textbf {\bibinfo {volume}
  {20}},\ \bibinfo {pages} {61} (\bibinfo {year} {2009})}\BibitemShut {NoStop}%
\bibitem [{\citenamefont {Gilmer}\ \emph {et~al.}(2017)\citenamefont {Gilmer},
  \citenamefont {Schoenholz}, \citenamefont {Riley}, \citenamefont {Vinyals},\
  and\ \citenamefont {Dahl}}]{gilmer2017}%
  \BibitemOpen
  \bibfield  {author} {\bibinfo {author} {\bibfnamefont {J.}~\bibnamefont
  {Gilmer}}, \bibinfo {author} {\bibfnamefont {S.~S.}\ \bibnamefont
  {Schoenholz}}, \bibinfo {author} {\bibfnamefont {P.~F.}\ \bibnamefont
  {Riley}}, \bibinfo {author} {\bibfnamefont {O.}~\bibnamefont {Vinyals}},\
  and\ \bibinfo {author} {\bibfnamefont {G.~E.}\ \bibnamefont {Dahl}},\
  }\bibfield  {title} {\bibinfo {title} {Neural message passing for quantum
  chemistry},\ }in\ \href {https://doi.org/10.48550/arXiv.1704.01212} {\emph
  {\bibinfo {booktitle} {Proceedings of the 34th International Conference on
  Machine Learning}}},\ Vol.~\bibinfo {volume} {70}\ (\bibinfo  {publisher}
  {PMLR},\ \bibinfo {year} {2017})\ pp.\ \bibinfo {pages} {1263--1272},\
  \Eprint {https://arxiv.org/abs/1704.01212} {arXiv:1704.01212} \BibitemShut
  {NoStop}%
\bibitem [{\citenamefont {Hamilton}\ \emph {et~al.}(2017)\citenamefont
  {Hamilton}, \citenamefont {Ying},\ and\ \citenamefont
  {Leskovec}}]{hamilton2017}%
  \BibitemOpen
  \bibfield  {author} {\bibinfo {author} {\bibfnamefont {W.~L.}\ \bibnamefont
  {Hamilton}}, \bibinfo {author} {\bibfnamefont {R.}~\bibnamefont {Ying}},\
  and\ \bibinfo {author} {\bibfnamefont {J.}~\bibnamefont {Leskovec}},\
  }\bibfield  {title} {\bibinfo {title} {Inductive representation learning on
  large graphs},\ }in\ \href {https://doi.org/10.48550/arXiv.1706.02216} {\emph
  {\bibinfo {booktitle} {Advances in Neural Information Processing Systems}}},\
  Vol.~\bibinfo {volume} {30}\ (\bibinfo {year} {2017})\ \Eprint
  {https://arxiv.org/abs/1706.02216} {arXiv:1706.02216} \BibitemShut {NoStop}%
\bibitem [{\citenamefont {Xu}\ \emph {et~al.}(2019)\citenamefont {Xu},
  \citenamefont {Hu}, \citenamefont {Leskovec},\ and\ \citenamefont
  {Jegelka}}]{xu2019}%
  \BibitemOpen
  \bibfield  {author} {\bibinfo {author} {\bibfnamefont {K.}~\bibnamefont
  {Xu}}, \bibinfo {author} {\bibfnamefont {W.}~\bibnamefont {Hu}}, \bibinfo
  {author} {\bibfnamefont {J.}~\bibnamefont {Leskovec}},\ and\ \bibinfo
  {author} {\bibfnamefont {S.}~\bibnamefont {Jegelka}},\ }\bibfield  {title}
  {\bibinfo {title} {How powerful are graph neural networks?},\ }in\ \href
  {https://doi.org/10.48550/arXiv.1810.00826} {\emph {\bibinfo {booktitle}
  {International Conference on Learning Representations (ICLR)}}}\ (\bibinfo
  {year} {2019})\ \Eprint {https://arxiv.org/abs/1810.00826} {arXiv:1810.00826}
  \BibitemShut {NoStop}%
\bibitem [{\citenamefont {Maron}\ \emph {et~al.}(2019)\citenamefont {Maron},
  \citenamefont {Ben-Hamu}, \citenamefont {Shamir},\ and\ \citenamefont
  {Lipman}}]{maron2019}%
  \BibitemOpen
  \bibfield  {author} {\bibinfo {author} {\bibfnamefont {H.}~\bibnamefont
  {Maron}}, \bibinfo {author} {\bibfnamefont {H.}~\bibnamefont {Ben-Hamu}},
  \bibinfo {author} {\bibfnamefont {N.}~\bibnamefont {Shamir}},\ and\ \bibinfo
  {author} {\bibfnamefont {Y.}~\bibnamefont {Lipman}},\ }\bibfield  {title}
  {\bibinfo {title} {Invariant and equivariant graph networks},\ }in\ \href
  {https://doi.org/10.48550/arXiv.1812.09902} {\emph {\bibinfo {booktitle}
  {International Conference on Learning Representations (ICLR)}}}\ (\bibinfo
  {year} {2019})\ \Eprint {https://arxiv.org/abs/1812.09902} {arXiv:1812.09902}
  \BibitemShut {NoStop}%
\bibitem [{\citenamefont {Xie}\ and\ \citenamefont {Grossman}(2018)}]{xie2018}%
  \BibitemOpen
  \bibfield  {author} {\bibinfo {author} {\bibfnamefont {T.}~\bibnamefont
  {Xie}}\ and\ \bibinfo {author} {\bibfnamefont {J.~C.}\ \bibnamefont
  {Grossman}},\ }\bibfield  {title} {\bibinfo {title} {Crystal graph
  convolutional neural networks for an accurate and interpretable prediction of
  material properties},\ }\href
  {https://doi.org/10.1103/PhysRevLett.120.145301} {\bibfield  {journal}
  {\bibinfo  {journal} {Phys. Rev. Lett.}\ }\textbf {\bibinfo {volume} {120}},\
  \bibinfo {pages} {145301} (\bibinfo {year} {2018})}\BibitemShut {NoStop}%
\bibitem [{\citenamefont {Sch\"utt}\ \emph {et~al.}(2018)\citenamefont
  {Sch\"utt}, \citenamefont {Sauceda}, \citenamefont {Kindermans},
  \citenamefont {Tkatchenko},\ and\ \citenamefont {M\"uller}}]{schutt2018}%
  \BibitemOpen
  \bibfield  {author} {\bibinfo {author} {\bibfnamefont {K.~T.}\ \bibnamefont
  {Sch\"utt}}, \bibinfo {author} {\bibfnamefont {H.~E.}\ \bibnamefont
  {Sauceda}}, \bibinfo {author} {\bibfnamefont {P.-J.}\ \bibnamefont
  {Kindermans}}, \bibinfo {author} {\bibfnamefont {A.}~\bibnamefont
  {Tkatchenko}},\ and\ \bibinfo {author} {\bibfnamefont {K.-R.}\ \bibnamefont
  {M\"uller}},\ }\bibfield  {title} {\bibinfo {title} {Schnet – a deep
  learning architecture for molecules and materials},\ }\href
  {https://doi.org/10.1063/1.5019779} {\bibfield  {journal} {\bibinfo
  {journal} {The Journal of Chemical Physics}\ }\textbf {\bibinfo {volume}
  {148}},\ \bibinfo {pages} {241722} (\bibinfo {year} {2018})}\BibitemShut
  {NoStop}%
\bibitem [{\citenamefont {Choudhary}\ and\ \citenamefont
  {DeCost}(2021)}]{choudhary2021}%
  \BibitemOpen
  \bibfield  {author} {\bibinfo {author} {\bibfnamefont {K.}~\bibnamefont
  {Choudhary}}\ and\ \bibinfo {author} {\bibfnamefont {B.}~\bibnamefont
  {DeCost}},\ }\bibfield  {title} {\bibinfo {title} {Atomistic line graph
  neural network for improved materials property predictions},\ }\href
  {https://doi.org/10.1038/s41524-021-00650-1} {\bibfield  {journal} {\bibinfo
  {journal} {npj Computational Materials}\ }\textbf {\bibinfo {volume} {7}},\
  \bibinfo {pages} {185} (\bibinfo {year} {2021})}\BibitemShut {NoStop}%
\bibitem [{\citenamefont {Dai}\ \emph {et~al.}(2021)\citenamefont {Dai},
  \citenamefont {Demirel}, \citenamefont {Liang},\ and\ \citenamefont
  {Hu}}]{dai2021}%
  \BibitemOpen
  \bibfield  {author} {\bibinfo {author} {\bibfnamefont {M.}~\bibnamefont
  {Dai}}, \bibinfo {author} {\bibfnamefont {M.~F.}\ \bibnamefont {Demirel}},
  \bibinfo {author} {\bibfnamefont {Y.}~\bibnamefont {Liang}},\ and\ \bibinfo
  {author} {\bibfnamefont {J.-M.}\ \bibnamefont {Hu}},\ }\bibfield  {title}
  {\bibinfo {title} {Graph neural networks for an accurate and interpretable
  prediction of the properties of polycrystalline materials},\ }\href
  {https://doi.org/10.1038/s41524-021-00574-w} {\bibfield  {journal} {\bibinfo
  {journal} {npj Computational Materials}\ }\textbf {\bibinfo {volume} {7}},\
  \bibinfo {pages} {103} (\bibinfo {year} {2021})}\BibitemShut {NoStop}%
\bibitem [{\citenamefont {Reiser}\ \emph {et~al.}(2022)\citenamefont {Reiser},
  \citenamefont {Neubert}, \citenamefont {Eberhard}, \citenamefont {Torresi},
  \citenamefont {Zhou}, \citenamefont {Shao}, \citenamefont {Metni},
  \citenamefont {van Hoesel}, \citenamefont {Schopmans}, \citenamefont
  {Sommer},\ and\ \citenamefont {Friederich}}]{reiser2022}%
  \BibitemOpen
  \bibfield  {author} {\bibinfo {author} {\bibfnamefont {P.}~\bibnamefont
  {Reiser}}, \bibinfo {author} {\bibfnamefont {M.}~\bibnamefont {Neubert}},
  \bibinfo {author} {\bibfnamefont {A.}~\bibnamefont {Eberhard}}, \bibinfo
  {author} {\bibfnamefont {L.}~\bibnamefont {Torresi}}, \bibinfo {author}
  {\bibfnamefont {C.}~\bibnamefont {Zhou}}, \bibinfo {author} {\bibfnamefont
  {C.}~\bibnamefont {Shao}}, \bibinfo {author} {\bibfnamefont {H.}~\bibnamefont
  {Metni}}, \bibinfo {author} {\bibfnamefont {C.}~\bibnamefont {van Hoesel}},
  \bibinfo {author} {\bibfnamefont {H.}~\bibnamefont {Schopmans}}, \bibinfo
  {author} {\bibfnamefont {T.}~\bibnamefont {Sommer}},\ and\ \bibinfo {author}
  {\bibfnamefont {P.}~\bibnamefont {Friederich}},\ }\bibfield  {title}
  {\bibinfo {title} {Graph neural networks for materials science and
  chemistry},\ }\href {https://doi.org/10.1038/s43246-022-00315-6} {\bibfield
  {journal} {\bibinfo  {journal} {Communications Materials}\ }\textbf {\bibinfo
  {volume} {3}},\ \bibinfo {pages} {93} (\bibinfo {year} {2022})}\BibitemShut
  {NoStop}%
\bibitem [{\citenamefont {Batatia}\ \emph {et~al.}(2022)\citenamefont
  {Batatia}, \citenamefont {Kovács}, \citenamefont {Simm}, \citenamefont
  {Ortner},\ and\ \citenamefont {Cs\'anyi}}]{batatia2022}%
  \BibitemOpen
  \bibfield  {author} {\bibinfo {author} {\bibfnamefont {I.}~\bibnamefont
  {Batatia}}, \bibinfo {author} {\bibfnamefont {D.~P.}\ \bibnamefont
  {Kovács}}, \bibinfo {author} {\bibfnamefont {G.~N.~C.}\ \bibnamefont
  {Simm}}, \bibinfo {author} {\bibfnamefont {C.}~\bibnamefont {Ortner}},\ and\
  \bibinfo {author} {\bibfnamefont {G.}~\bibnamefont {Cs\'anyi}},\ }\bibfield
  {title} {\bibinfo {title} {Mace: Higher order equivariant message passing
  neural networks for fast and accurate force fields},\ }in\ \href
  {https://proceedings.neurips.cc/paper/2022/hash/4a36c3c51af11ed9f34615b81edb5bbc-Abstract-Conference.html}
  {\emph {\bibinfo {booktitle} {Advances in Neural Information Processing
  Systems}}},\ Vol.~\bibinfo {volume} {35}\ (\bibinfo {year} {2022})\ pp.\
  \bibinfo {pages} {11423--11436},\ \Eprint {https://arxiv.org/abs/2206.07697}
  {arXiv:2206.07697 [stat.ML]} \BibitemShut {NoStop}%
\bibitem [{\citenamefont {Cohen}\ and\ \citenamefont
  {Welling}(2016)}]{cohen2016}%
  \BibitemOpen
  \bibfield  {author} {\bibinfo {author} {\bibfnamefont {T.~S.}\ \bibnamefont
  {Cohen}}\ and\ \bibinfo {author} {\bibfnamefont {M.}~\bibnamefont
  {Welling}},\ }\bibfield  {title} {\bibinfo {title} {Group equivariant
  convolutional networks},\ }in\ \href
  {https://proceedings.mlr.press/v48/cohenc16.html} {\emph {\bibinfo
  {booktitle} {Proceedings of the 33rd International Conference on Machine
  Learning}}},\ \bibinfo {series} {Proceedings of Machine Learning Research},
  Vol.~\bibinfo {volume} {48}\ (\bibinfo  {publisher} {PMLR},\ \bibinfo {year}
  {2016})\ pp.\ \bibinfo {pages} {2990--2999},\ \Eprint
  {https://arxiv.org/abs/1602.07576} {arXiv:1602.07576 [cs.LG]} \BibitemShut
  {NoStop}%
\bibitem [{\citenamefont {Cohen}\ \emph {et~al.}(2018)\citenamefont {Cohen},
  \citenamefont {Geiger}, \citenamefont {Weiler},\ and\ \citenamefont
  {Welling}}]{cohen2018}%
  \BibitemOpen
  \bibfield  {author} {\bibinfo {author} {\bibfnamefont {T.~S.}\ \bibnamefont
  {Cohen}}, \bibinfo {author} {\bibfnamefont {M.}~\bibnamefont {Geiger}},
  \bibinfo {author} {\bibfnamefont {M.}~\bibnamefont {Weiler}},\ and\ \bibinfo
  {author} {\bibfnamefont {M.}~\bibnamefont {Welling}},\ }\bibfield  {title}
  {\bibinfo {title} {A general theory of equivariant cnns on homogeneous
  spaces},\ }in\ \href
  {https://proceedings.neurips.cc/paper/2018/hash/488e4104520c6aab692863cc1dba45af-Abstract.html}
  {\emph {\bibinfo {booktitle} {Advances in Neural Information Processing
  Systems}}},\ Vol.~\bibinfo {volume} {31}\ (\bibinfo {year} {2018})\ \Eprint
  {https://arxiv.org/abs/1811.02017} {arXiv:1811.02017 [cs.LG]} \BibitemShut
  {NoStop}%
\bibitem [{\citenamefont {Weiler}\ \emph {et~al.}(2018)\citenamefont {Weiler},
  \citenamefont {Geiger}, \citenamefont {Welling}, \citenamefont {Boomsma},\
  and\ \citenamefont {Cohen}}]{weiler2018}%
  \BibitemOpen
  \bibfield  {author} {\bibinfo {author} {\bibfnamefont {M.}~\bibnamefont
  {Weiler}}, \bibinfo {author} {\bibfnamefont {M.}~\bibnamefont {Geiger}},
  \bibinfo {author} {\bibfnamefont {M.}~\bibnamefont {Welling}}, \bibinfo
  {author} {\bibfnamefont {W.}~\bibnamefont {Boomsma}},\ and\ \bibinfo {author}
  {\bibfnamefont {T.~S.}\ \bibnamefont {Cohen}},\ }\bibfield  {title} {\bibinfo
  {title} {3d steerable cnns: Learning rotationally equivariant features in
  volumetric data},\ }in\ \href
  {https://proceedings.neurips.cc/paper/2018/hash/488e4104520c6aab692863cc1dba45af-Abstract.html}
  {\emph {\bibinfo {booktitle} {Advances in Neural Information Processing
  Systems}}},\ Vol.~\bibinfo {volume} {31}\ (\bibinfo  {publisher} {Curran
  Associates, Inc.},\ \bibinfo {year} {2018})\ \Eprint
  {https://arxiv.org/abs/1807.02547} {arXiv:1807.02547 [cs.LG]} \BibitemShut
  {NoStop}%
\bibitem [{\citenamefont {Kondor}(2025)}]{kondor2025}%
  \BibitemOpen
  \bibfield  {author} {\bibinfo {author} {\bibfnamefont {R.}~\bibnamefont
  {Kondor}},\ }\bibfield  {title} {\bibinfo {title} {The principles behind
  equivariant neural networks for physics and chemistry},\ }\href
  {https://doi.org/10.1073/pnas.2415656122} {\bibfield  {journal} {\bibinfo
  {journal} {Proceedings of the National Academy of Sciences}\ }\textbf
  {\bibinfo {volume} {122}},\ \bibinfo {pages} {e2415656122} (\bibinfo {year}
  {2025})}\BibitemShut {NoStop}%
\bibitem [{\citenamefont {Batzner}\ \emph {et~al.}(2022)\citenamefont
  {Batzner}, \citenamefont {Musaelian}, \citenamefont {Sun}, \citenamefont
  {Geiger}, \citenamefont {Mailoa}, \citenamefont {Kornbluth}, \citenamefont
  {Molinari}, \citenamefont {Smidt},\ and\ \citenamefont
  {Kozinsky}}]{batzner2022}%
  \BibitemOpen
  \bibfield  {author} {\bibinfo {author} {\bibfnamefont {S.}~\bibnamefont
  {Batzner}}, \bibinfo {author} {\bibfnamefont {A.}~\bibnamefont {Musaelian}},
  \bibinfo {author} {\bibfnamefont {L.}~\bibnamefont {Sun}}, \bibinfo {author}
  {\bibfnamefont {M.}~\bibnamefont {Geiger}}, \bibinfo {author} {\bibfnamefont
  {J.~P.}\ \bibnamefont {Mailoa}}, \bibinfo {author} {\bibfnamefont
  {M.}~\bibnamefont {Kornbluth}}, \bibinfo {author} {\bibfnamefont
  {N.}~\bibnamefont {Molinari}}, \bibinfo {author} {\bibfnamefont {T.~E.}\
  \bibnamefont {Smidt}},\ and\ \bibinfo {author} {\bibfnamefont
  {B.}~\bibnamefont {Kozinsky}},\ }\bibfield  {title} {\bibinfo {title}
  {E(3)-equivariant graph neural networks for data-efficient and accurate
  interatomic potentials},\ }\href {https://doi.org/10.1038/s41467-022-29939-5}
  {\bibfield  {journal} {\bibinfo  {journal} {Nature Communications}\ }\textbf
  {\bibinfo {volume} {13}},\ \bibinfo {pages} {2453} (\bibinfo {year}
  {2022})}\BibitemShut {NoStop}%
\bibitem [{\citenamefont {Kaba}\ and\ \citenamefont
  {Ravanbakhsh}(2022)}]{kaba2022}%
  \BibitemOpen
  \bibfield  {author} {\bibinfo {author} {\bibfnamefont {S.-O.}\ \bibnamefont
  {Kaba}}\ and\ \bibinfo {author} {\bibfnamefont {S.}~\bibnamefont
  {Ravanbakhsh}},\ }\bibfield  {title} {\bibinfo {title} {Equivariant networks
  for crystal structures},\ }in\ \href
  {https://proceedings.neurips.cc/paper/2022/hash/1abed6ee581b9ceb4e2ddf37822c7fcb-Abstract-Conference.html}
  {\emph {\bibinfo {booktitle} {Advances in Neural Information Processing
  Systems}}},\ Vol.~\bibinfo {volume} {35}\ (\bibinfo {year} {2022})\ pp.\
  \bibinfo {pages} {300--314},\ \Eprint {https://arxiv.org/abs/2211.15420}
  {arXiv:2211.15420 [cs.LG]} \BibitemShut {NoStop}%
\bibitem [{\citenamefont {Musaelian}\ \emph {et~al.}(2023)\citenamefont
  {Musaelian}, \citenamefont {Batzner}, \citenamefont {Johansson},
  \citenamefont {Sun}, \citenamefont {Owen}, \citenamefont {Kornbluth},\ and\
  \citenamefont {Kozinsky}}]{musaelian2023}%
  \BibitemOpen
  \bibfield  {author} {\bibinfo {author} {\bibfnamefont {A.}~\bibnamefont
  {Musaelian}}, \bibinfo {author} {\bibfnamefont {S.}~\bibnamefont {Batzner}},
  \bibinfo {author} {\bibfnamefont {A.}~\bibnamefont {Johansson}}, \bibinfo
  {author} {\bibfnamefont {L.}~\bibnamefont {Sun}}, \bibinfo {author}
  {\bibfnamefont {C.~J.}\ \bibnamefont {Owen}}, \bibinfo {author}
  {\bibfnamefont {M.}~\bibnamefont {Kornbluth}},\ and\ \bibinfo {author}
  {\bibfnamefont {B.}~\bibnamefont {Kozinsky}},\ }\bibfield  {title} {\bibinfo
  {title} {Learning local equivariant representations for large-scale atomistic
  dynamics},\ }\href {https://doi.org/10.1038/s41467-023-36329-y} {\bibfield
  {journal} {\bibinfo  {journal} {Nature Communications}\ }\textbf {\bibinfo
  {volume} {14}},\ \bibinfo {pages} {579} (\bibinfo {year} {2023})}\BibitemShut
  {NoStop}%
\bibitem [{\citenamefont {Gong}\ \emph {et~al.}(2023)\citenamefont {Gong},
  \citenamefont {Li}, \citenamefont {Zou}, \citenamefont {Xu}, \citenamefont
  {Duan},\ and\ \citenamefont {Xu}}]{gong2023}%
  \BibitemOpen
  \bibfield  {author} {\bibinfo {author} {\bibfnamefont {X.}~\bibnamefont
  {Gong}}, \bibinfo {author} {\bibfnamefont {H.}~\bibnamefont {Li}}, \bibinfo
  {author} {\bibfnamefont {N.}~\bibnamefont {Zou}}, \bibinfo {author}
  {\bibfnamefont {R.}~\bibnamefont {Xu}}, \bibinfo {author} {\bibfnamefont
  {W.}~\bibnamefont {Duan}},\ and\ \bibinfo {author} {\bibfnamefont
  {Y.}~\bibnamefont {Xu}},\ }\bibfield  {title} {\bibinfo {title} {General
  framework for e(3)-equivariant neural network representation of density
  functional theory hamiltonian},\ }\href
  {https://doi.org/10.1038/s41467-023-38468-8} {\bibfield  {journal} {\bibinfo
  {journal} {Nature Communications}\ }\textbf {\bibinfo {volume} {14}},\
  \bibinfo {pages} {2848} (\bibinfo {year} {2023})}\BibitemShut {NoStop}%
\bibitem [{\citenamefont {Batatia}\ \emph {et~al.}(2025)\citenamefont
  {Batatia}, \citenamefont {Batzner}, \citenamefont {Kov{\'a}cs}, \citenamefont
  {Musaelian}, \citenamefont {Simm}, \citenamefont {Drautz}, \citenamefont
  {Ortner}, \citenamefont {Kozinsky},\ and\ \citenamefont
  {Cs{\'a}nyi}}]{batatia2025}%
  \BibitemOpen
  \bibfield  {author} {\bibinfo {author} {\bibfnamefont {I.}~\bibnamefont
  {Batatia}}, \bibinfo {author} {\bibfnamefont {S.}~\bibnamefont {Batzner}},
  \bibinfo {author} {\bibfnamefont {D.~P.}\ \bibnamefont {Kov{\'a}cs}},
  \bibinfo {author} {\bibfnamefont {A.}~\bibnamefont {Musaelian}}, \bibinfo
  {author} {\bibfnamefont {G.~N.~C.}\ \bibnamefont {Simm}}, \bibinfo {author}
  {\bibfnamefont {R.}~\bibnamefont {Drautz}}, \bibinfo {author} {\bibfnamefont
  {C.}~\bibnamefont {Ortner}}, \bibinfo {author} {\bibfnamefont
  {B.}~\bibnamefont {Kozinsky}},\ and\ \bibinfo {author} {\bibfnamefont
  {G.}~\bibnamefont {Cs{\'a}nyi}},\ }\bibfield  {title} {\bibinfo {title} {The
  design space of e(3)-equivariant atom-centred interatomic potentials},\
  }\href {https://doi.org/10.1038/s42256-024-00956-x} {\bibfield  {journal}
  {\bibinfo  {journal} {Nature Machine Intelligence}\ }\textbf {\bibinfo
  {volume} {7}},\ \bibinfo {pages} {56} (\bibinfo {year} {2025})}\BibitemShut
  {NoStop}%
\bibitem [{\citenamefont {Yang}\ \emph {et~al.}(2025)\citenamefont {Yang},
  \citenamefont {Wang}, \citenamefont {Li}, \citenamefont {Lv}, \citenamefont
  {Chen},\ and\ \citenamefont {Shen}}]{yang2025}%
  \BibitemOpen
  \bibfield  {author} {\bibinfo {author} {\bibfnamefont {Z.}~\bibnamefont
  {Yang}}, \bibinfo {author} {\bibfnamefont {X.}~\bibnamefont {Wang}}, \bibinfo
  {author} {\bibfnamefont {Y.}~\bibnamefont {Li}}, \bibinfo {author}
  {\bibfnamefont {Q.}~\bibnamefont {Lv}}, \bibinfo {author} {\bibfnamefont
  {C.~Y.-C.}\ \bibnamefont {Chen}},\ and\ \bibinfo {author} {\bibfnamefont
  {L.}~\bibnamefont {Shen}},\ }\bibfield  {title} {\bibinfo {title} {Efficient
  equivariant model for machine learning interatomic potentials},\ }\href
  {https://doi.org/10.1038/s41524-025-01535-3} {\bibfield  {journal} {\bibinfo
  {journal} {npj Computational Materials}\ }\textbf {\bibinfo {volume} {11}},\
  \bibinfo {pages} {49} (\bibinfo {year} {2025})}\BibitemShut {NoStop}%
\bibitem [{\citenamefont {Nebgen}\ \emph {et~al.}(2018)\citenamefont {Nebgen},
  \citenamefont {Lubbers}, \citenamefont {Smith}, \citenamefont {Sifain},
  \citenamefont {Lokhov}, \citenamefont {Isayev}, \citenamefont {Roitberg},
  \citenamefont {Barros},\ and\ \citenamefont {Tretiak}}]{Nebgen2018}%
  \BibitemOpen
  \bibfield  {author} {\bibinfo {author} {\bibfnamefont {B.}~\bibnamefont
  {Nebgen}}, \bibinfo {author} {\bibfnamefont {N.}~\bibnamefont {Lubbers}},
  \bibinfo {author} {\bibfnamefont {J.~S.}\ \bibnamefont {Smith}}, \bibinfo
  {author} {\bibfnamefont {A.~E.}\ \bibnamefont {Sifain}}, \bibinfo {author}
  {\bibfnamefont {A.}~\bibnamefont {Lokhov}}, \bibinfo {author} {\bibfnamefont
  {O.}~\bibnamefont {Isayev}}, \bibinfo {author} {\bibfnamefont {A.~E.}\
  \bibnamefont {Roitberg}}, \bibinfo {author} {\bibfnamefont {K.}~\bibnamefont
  {Barros}},\ and\ \bibinfo {author} {\bibfnamefont {S.}~\bibnamefont
  {Tretiak}},\ }\bibfield  {title} {\bibinfo {title} {Transferable dynamic
  molecular charge assignment using deep neural networks},\ }\href
  {https://doi.org/10.1021/acs.jctc.8b00524} {\bibfield  {journal} {\bibinfo
  {journal} {Journal of Chemical Theory and Computation}\ }\textbf {\bibinfo
  {volume} {14}},\ \bibinfo {pages} {4687} (\bibinfo {year}
  {2018})}\BibitemShut {NoStop}%
\bibitem [{\citenamefont {Smith}\ \emph {et~al.}(2019)\citenamefont {Smith},
  \citenamefont {Isayev},\ and\ \citenamefont {Roitberg}}]{Smith2019}%
  \BibitemOpen
  \bibfield  {author} {\bibinfo {author} {\bibfnamefont {J.~S.}\ \bibnamefont
  {Smith}}, \bibinfo {author} {\bibfnamefont {O.}~\bibnamefont {Isayev}},\ and\
  \bibinfo {author} {\bibfnamefont {A.~E.}\ \bibnamefont {Roitberg}},\
  }\bibfield  {title} {\bibinfo {title} {Approaching coupled cluster accuracy
  with a general-purpose neural network potential through transfer learning},\
  }\href {https://doi.org/10.1038/s41467-019-10827-4} {\bibfield  {journal}
  {\bibinfo  {journal} {Nature Communications}\ }\textbf {\bibinfo {volume}
  {10}},\ \bibinfo {pages} {2903} (\bibinfo {year} {2019})}\BibitemShut
  {NoStop}%
\bibitem [{\citenamefont {Zhou}\ \emph {et~al.}(2022)\citenamefont {Zhou},
  \citenamefont {Nebgen}, \citenamefont {Lubbers}, \citenamefont {Malone},
  \citenamefont {Niklasson},\ and\ \citenamefont {Tretiak}}]{Zhou2022}%
  \BibitemOpen
  \bibfield  {author} {\bibinfo {author} {\bibfnamefont {G.}~\bibnamefont
  {Zhou}}, \bibinfo {author} {\bibfnamefont {B.}~\bibnamefont {Nebgen}},
  \bibinfo {author} {\bibfnamefont {N.}~\bibnamefont {Lubbers}}, \bibinfo
  {author} {\bibfnamefont {F.}~\bibnamefont {Malone}}, \bibinfo {author}
  {\bibfnamefont {A.}~\bibnamefont {Niklasson}},\ and\ \bibinfo {author}
  {\bibfnamefont {S.}~\bibnamefont {Tretiak}},\ }\bibfield  {title} {\bibinfo
  {title} {Deep learning of dynamically responsive chemical hamiltonians with
  semiempirical quantum mechanics},\ }\href
  {https://doi.org/10.1073/pnas.2120333119} {\bibfield  {journal} {\bibinfo
  {journal} {Proceedings of the National Academy of Sciences}\ }\textbf
  {\bibinfo {volume} {119}},\ \bibinfo {pages} {e2120333119} (\bibinfo {year}
  {2022})}\BibitemShut {NoStop}%
\bibitem [{\citenamefont {Chigaev}\ \emph {et~al.}(2023)\citenamefont
  {Chigaev}, \citenamefont {Smith}, \citenamefont {Anaya}, \citenamefont
  {Nebgen}, \citenamefont {Bettencourt}, \citenamefont {Barros},\ and\
  \citenamefont {Lubbers}}]{Chigaev2023}%
  \BibitemOpen
  \bibfield  {author} {\bibinfo {author} {\bibfnamefont {M.}~\bibnamefont
  {Chigaev}}, \bibinfo {author} {\bibfnamefont {J.~S.}\ \bibnamefont {Smith}},
  \bibinfo {author} {\bibfnamefont {S.}~\bibnamefont {Anaya}}, \bibinfo
  {author} {\bibfnamefont {B.}~\bibnamefont {Nebgen}}, \bibinfo {author}
  {\bibfnamefont {M.}~\bibnamefont {Bettencourt}}, \bibinfo {author}
  {\bibfnamefont {K.}~\bibnamefont {Barros}},\ and\ \bibinfo {author}
  {\bibfnamefont {N.}~\bibnamefont {Lubbers}},\ }\bibfield  {title} {\bibinfo
  {title} {Lightweight and effective tensor sensitivity for atomistic neural
  networks},\ }\href {https://doi.org/10.1063/5.0142127} {\bibfield  {journal}
  {\bibinfo  {journal} {The Journal of Chemical Physics}\ }\textbf {\bibinfo
  {volume} {158}},\ \bibinfo {pages} {184108} (\bibinfo {year}
  {2023})}\BibitemShut {NoStop}%
\bibitem [{\citenamefont {Zhang}\ \emph {et~al.}(2025)\citenamefont {Zhang},
  \citenamefont {Chigaev}, \citenamefont {Isayev}, \citenamefont {Messerly},\
  and\ \citenamefont {Lubbers}}]{Zhang2025}%
  \BibitemOpen
  \bibfield  {author} {\bibinfo {author} {\bibfnamefont {S.}~\bibnamefont
  {Zhang}}, \bibinfo {author} {\bibfnamefont {M.}~\bibnamefont {Chigaev}},
  \bibinfo {author} {\bibfnamefont {O.}~\bibnamefont {Isayev}}, \bibinfo
  {author} {\bibfnamefont {R.~A.}\ \bibnamefont {Messerly}},\ and\ \bibinfo
  {author} {\bibfnamefont {N.}~\bibnamefont {Lubbers}},\ }\bibfield  {title}
  {\bibinfo {title} {Including physics-informed atomization constraints in
  neural networks for reactive chemistry},\ }\href
  {https://doi.org/10.1021/acs.jcim.5c00341} {\bibfield  {journal} {\bibinfo
  {journal} {Journal of Chemical Information and Modeling}\ }\textbf {\bibinfo
  {volume} {65}},\ \bibinfo {pages} {4367} (\bibinfo {year}
  {2025})}\BibitemShut {NoStop}%
\bibitem [{\citenamefont {Eckhoff}\ and\ \citenamefont
  {Behler}(2021)}]{Eckhoff2021}%
  \BibitemOpen
  \bibfield  {author} {\bibinfo {author} {\bibfnamefont {M.}~\bibnamefont
  {Eckhoff}}\ and\ \bibinfo {author} {\bibfnamefont {J.}~\bibnamefont
  {Behler}},\ }\bibfield  {title} {\bibinfo {title} {High-dimensional neural
  network potentials for magnetic systems using spin-dependent atom-centered
  symmetry functions},\ }\href {https://doi.org/10.1038/s41524-021-00636-z}
  {\bibfield  {journal} {\bibinfo  {journal} {npj Computational Materials}\
  }\textbf {\bibinfo {volume} {7}},\ \bibinfo {pages} {170} (\bibinfo {year}
  {2021})}\BibitemShut {NoStop}%
\bibitem [{\citenamefont {Kotykhov}\ \emph {et~al.}(2024)\citenamefont
  {Kotykhov}, \citenamefont {Gubaev}, \citenamefont {Sotskov}, \citenamefont
  {Tantardini}, \citenamefont {Hodapp}, \citenamefont {Shapeev},\ and\
  \citenamefont {Novikov}}]{Kotykhov2024}%
  \BibitemOpen
  \bibfield  {author} {\bibinfo {author} {\bibfnamefont {A.~S.}\ \bibnamefont
  {Kotykhov}}, \bibinfo {author} {\bibfnamefont {K.}~\bibnamefont {Gubaev}},
  \bibinfo {author} {\bibfnamefont {V.}~\bibnamefont {Sotskov}}, \bibinfo
  {author} {\bibfnamefont {C.}~\bibnamefont {Tantardini}}, \bibinfo {author}
  {\bibfnamefont {M.}~\bibnamefont {Hodapp}}, \bibinfo {author} {\bibfnamefont
  {A.~V.}\ \bibnamefont {Shapeev}},\ and\ \bibinfo {author} {\bibfnamefont
  {I.~S.}\ \bibnamefont {Novikov}},\ }\bibfield  {title} {\bibinfo {title}
  {Fitting to magnetic forces improves the reliability of magnetic moment
  tensor potentials},\ }\href {https://doi.org/10.1016/j.commatsci.2024.113331}
  {\bibfield  {journal} {\bibinfo  {journal} {Computational Materials Science}\
  }\textbf {\bibinfo {volume} {245}},\ \bibinfo {pages} {113331} (\bibinfo
  {year} {2024})}\BibitemShut {NoStop}%
\bibitem [{\citenamefont {Yu}\ \emph {et~al.}(2024{\natexlab{a}})\citenamefont
  {Yu}, \citenamefont {Zhong}, \citenamefont {Hong}, \citenamefont {Xu},
  \citenamefont {Ren}, \citenamefont {Gong},\ and\ \citenamefont
  {Xiang}}]{Yu2024}%
  \BibitemOpen
  \bibfield  {author} {\bibinfo {author} {\bibfnamefont {H.}~\bibnamefont
  {Yu}}, \bibinfo {author} {\bibfnamefont {Y.}~\bibnamefont {Zhong}}, \bibinfo
  {author} {\bibfnamefont {L.}~\bibnamefont {Hong}}, \bibinfo {author}
  {\bibfnamefont {C.}~\bibnamefont {Xu}}, \bibinfo {author} {\bibfnamefont
  {W.}~\bibnamefont {Ren}}, \bibinfo {author} {\bibfnamefont {X.}~\bibnamefont
  {Gong}},\ and\ \bibinfo {author} {\bibfnamefont {H.}~\bibnamefont {Xiang}},\
  }\bibfield  {title} {\bibinfo {title} {Spin-dependent graph neural network
  potential for magnetic materials},\ }\href
  {https://doi.org/10.1103/PhysRevB.109.144426} {\bibfield  {journal} {\bibinfo
   {journal} {Phys. Rev. B}\ }\textbf {\bibinfo {volume} {109}},\ \bibinfo
  {pages} {144426} (\bibinfo {year} {2024}{\natexlab{a}})}\BibitemShut
  {NoStop}%
\bibitem [{\citenamefont {Yu}\ \emph {et~al.}(2024{\natexlab{b}})\citenamefont
  {Yu}, \citenamefont {Liu}, \citenamefont {Zhong}, \citenamefont {Hong},
  \citenamefont {Ji}, \citenamefont {Xu}, \citenamefont {Gong},\ and\
  \citenamefont {Xiang}}]{Yu2024b}%
  \BibitemOpen
  \bibfield  {author} {\bibinfo {author} {\bibfnamefont {H.}~\bibnamefont
  {Yu}}, \bibinfo {author} {\bibfnamefont {B.}~\bibnamefont {Liu}}, \bibinfo
  {author} {\bibfnamefont {Y.}~\bibnamefont {Zhong}}, \bibinfo {author}
  {\bibfnamefont {L.}~\bibnamefont {Hong}}, \bibinfo {author} {\bibfnamefont
  {J.}~\bibnamefont {Ji}}, \bibinfo {author} {\bibfnamefont {C.}~\bibnamefont
  {Xu}}, \bibinfo {author} {\bibfnamefont {X.}~\bibnamefont {Gong}},\ and\
  \bibinfo {author} {\bibfnamefont {H.}~\bibnamefont {Xiang}},\ }\bibfield
  {title} {\bibinfo {title} {Physics-informed time-reversal equivariant neural
  network potential for magnetic materials},\ }\href
  {https://doi.org/10.1103/PhysRevB.110.104427} {\bibfield  {journal} {\bibinfo
   {journal} {Phys. Rev. B}\ }\textbf {\bibinfo {volume} {110}},\ \bibinfo
  {pages} {104427} (\bibinfo {year} {2024}{\natexlab{b}})}\BibitemShut
  {NoStop}%
\bibitem [{\citenamefont {Huang}\ \emph {et~al.}(2025)\citenamefont {Huang},
  \citenamefont {Yang}, \citenamefont {Liu},\ and\ \citenamefont
  {Xu}}]{Huang2025}%
  \BibitemOpen
  \bibfield  {author} {\bibinfo {author} {\bibfnamefont {Z.}~\bibnamefont
  {Huang}}, \bibinfo {author} {\bibfnamefont {T.}~\bibnamefont {Yang}},
  \bibinfo {author} {\bibfnamefont {H.}~\bibnamefont {Liu}},\ and\ \bibinfo
  {author} {\bibfnamefont {B.}~\bibnamefont {Xu}},\ }\bibfield  {title}
  {\bibinfo {title} {Simultaneous optimization of lattice and spin
  configurations in atomic scale simulation of magnetic materials},\ }\href
  {https://doi.org/10.1103/PhysRevB.111.134412} {\bibfield  {journal} {\bibinfo
   {journal} {Phys. Rev. B}\ }\textbf {\bibinfo {volume} {111}},\ \bibinfo
  {pages} {134412} (\bibinfo {year} {2025})}\BibitemShut {NoStop}%
\bibitem [{\citenamefont {Xu}\ \emph {et~al.}(2025)\citenamefont {Xu},
  \citenamefont {Sanspeur}, \citenamefont {Kolluru}, \citenamefont {Deng},
  \citenamefont {Harrington}, \citenamefont {Farrell}, \citenamefont {Reuter},\
  and\ \citenamefont {Kitchin}}]{Xu2025}%
  \BibitemOpen
  \bibfield  {author} {\bibinfo {author} {\bibfnamefont {W.}~\bibnamefont
  {Xu}}, \bibinfo {author} {\bibfnamefont {R.~Y.}\ \bibnamefont {Sanspeur}},
  \bibinfo {author} {\bibfnamefont {A.}~\bibnamefont {Kolluru}}, \bibinfo
  {author} {\bibfnamefont {B.}~\bibnamefont {Deng}}, \bibinfo {author}
  {\bibfnamefont {P.}~\bibnamefont {Harrington}}, \bibinfo {author}
  {\bibfnamefont {S.}~\bibnamefont {Farrell}}, \bibinfo {author} {\bibfnamefont
  {K.}~\bibnamefont {Reuter}},\ and\ \bibinfo {author} {\bibfnamefont {J.~R.}\
  \bibnamefont {Kitchin}},\ }\bibfield  {title} {\bibinfo {title}
  {Spin-informed universal graph neural networks for simulating magnetic
  ordering},\ }\href {https://doi.org/10.1073/pnas.2422973122} {\bibfield
  {journal} {\bibinfo  {journal} {Proceedings of the National Academy of
  Sciences}\ }\textbf {\bibinfo {volume} {122}},\ \bibinfo {pages}
  {e2422973122} (\bibinfo {year} {2025})}\BibitemShut {NoStop}%
\bibitem [{\citenamefont {Zhang}\ and\ \citenamefont {Chern}(2021)}]{zhang21}%
  \BibitemOpen
  \bibfield  {author} {\bibinfo {author} {\bibfnamefont {P.}~\bibnamefont
  {Zhang}}\ and\ \bibinfo {author} {\bibfnamefont {G.-W.}\ \bibnamefont
  {Chern}},\ }\bibfield  {title} {\bibinfo {title} {Arrested phase separation
  in double-exchange models: Large-scale simulation enabled by machine
  learning},\ }\href {https://doi.org/10.1103/PhysRevLett.127.146401}
  {\bibfield  {journal} {\bibinfo  {journal} {Phys. Rev. Lett.}\ }\textbf
  {\bibinfo {volume} {127}},\ \bibinfo {pages} {146401} (\bibinfo {year}
  {2021})}\BibitemShut {NoStop}%
\bibitem [{\citenamefont {Zhang}\ and\ \citenamefont {Chern}(2023)}]{zhang23}%
  \BibitemOpen
  \bibfield  {author} {\bibinfo {author} {\bibfnamefont {P.}~\bibnamefont
  {Zhang}}\ and\ \bibinfo {author} {\bibfnamefont {G.-W.}\ \bibnamefont
  {Chern}},\ }\bibfield  {title} {\bibinfo {title} {Machine learning
  nonequilibrium electron forces for spin dynamics of itinerant magnets},\
  }\href {https://doi.org/10.1038/s41524-023-00990-0} {\bibfield  {journal}
  {\bibinfo  {journal} {npj Computational Materials}\ }\textbf {\bibinfo
  {volume} {9}},\ \bibinfo {pages} {32} (\bibinfo {year} {2023})}\BibitemShut
  {NoStop}%
\bibitem [{\citenamefont {Cheng}\ \emph {et~al.}(2023)\citenamefont {Cheng},
  \citenamefont {Zhang}, \citenamefont {Nguyen}, \citenamefont {Azarfar},
  \citenamefont {Chern},\ and\ \citenamefont {Baek}}]{cheng23b}%
  \BibitemOpen
  \bibfield  {author} {\bibinfo {author} {\bibfnamefont {X.}~\bibnamefont
  {Cheng}}, \bibinfo {author} {\bibfnamefont {S.}~\bibnamefont {Zhang}},
  \bibinfo {author} {\bibfnamefont {P.~C.~H.}\ \bibnamefont {Nguyen}}, \bibinfo
  {author} {\bibfnamefont {S.}~\bibnamefont {Azarfar}}, \bibinfo {author}
  {\bibfnamefont {G.-W.}\ \bibnamefont {Chern}},\ and\ \bibinfo {author}
  {\bibfnamefont {S.~S.}\ \bibnamefont {Baek}},\ }\bibfield  {title} {\bibinfo
  {title} {Convolutional neural networks for large-scale dynamical modeling of
  itinerant magnets},\ }\href
  {https://doi.org/10.1103/PhysRevResearch.5.033188} {\bibfield  {journal}
  {\bibinfo  {journal} {Phys. Rev. Res.}\ }\textbf {\bibinfo {volume} {5}},\
  \bibinfo {pages} {033188} (\bibinfo {year} {2023})}\BibitemShut {NoStop}%
\bibitem [{\citenamefont {Fan}\ \emph {et~al.}(2024)\citenamefont {Fan},
  \citenamefont {Zhang},\ and\ \citenamefont {Chern}}]{Fan24}%
  \BibitemOpen
  \bibfield  {author} {\bibinfo {author} {\bibfnamefont {Y.}~\bibnamefont
  {Fan}}, \bibinfo {author} {\bibfnamefont {S.}~\bibnamefont {Zhang}},\ and\
  \bibinfo {author} {\bibfnamefont {G.-W.}\ \bibnamefont {Chern}},\ }\bibfield
  {title} {\bibinfo {title} {Coarsening of chiral domains in itinerant electron
  magnets: A machine learning force-field approach},\ }\href
  {https://doi.org/10.1103/PhysRevB.110.245105} {\bibfield  {journal} {\bibinfo
   {journal} {Phys. Rev. B}\ }\textbf {\bibinfo {volume} {110}},\ \bibinfo
  {pages} {245105} (\bibinfo {year} {2024})}\BibitemShut {NoStop}%
\bibitem [{\citenamefont {Tyberg}\ \emph {et~al.}(2025)\citenamefont {Tyberg},
  \citenamefont {Fan},\ and\ \citenamefont {Chern}}]{tyberg25}%
  \BibitemOpen
  \bibfield  {author} {\bibinfo {author} {\bibfnamefont {A.}~\bibnamefont
  {Tyberg}}, \bibinfo {author} {\bibfnamefont {Y.}~\bibnamefont {Fan}},\ and\
  \bibinfo {author} {\bibfnamefont {G.-W.}\ \bibnamefont {Chern}},\ }\bibfield
  {title} {\bibinfo {title} {Machine learning force field model for kinetic
  monte carlo simulations of itinerant ising magnets},\ }\href
  {https://doi.org/10.1103/d3zm-pbr1} {\bibfield  {journal} {\bibinfo
  {journal} {Phys. Rev. B}\ }\textbf {\bibinfo {volume} {111}},\ \bibinfo
  {pages} {235132} (\bibinfo {year} {2025})}\BibitemShut {NoStop}%
\bibitem [{\citenamefont {Chern}\ \emph
  {et~al.}(2026{\natexlab{a}})\citenamefont {Chern}, \citenamefont {Fan},
  \citenamefont {Zhang},\ and\ \citenamefont {Zhang}}]{Chern2026}%
  \BibitemOpen
  \bibfield  {author} {\bibinfo {author} {\bibfnamefont {G.-W.}\ \bibnamefont
  {Chern}}, \bibinfo {author} {\bibfnamefont {Y.}~\bibnamefont {Fan}}, \bibinfo
  {author} {\bibfnamefont {S.}~\bibnamefont {Zhang}},\ and\ \bibinfo {author}
  {\bibfnamefont {P.}~\bibnamefont {Zhang}},\ }\bibfield  {title} {\bibinfo
  {title} {Machine-learning force-field models for dynamical simulations of
  metallic magnets},\ }\href {https://doi.org/10.1063/9.0001056} {\bibfield
  {journal} {\bibinfo  {journal} {AIP Advances}\ }\textbf {\bibinfo {volume}
  {16}},\ \bibinfo {pages} {025031} (\bibinfo {year}
  {2026}{\natexlab{a}})}\BibitemShut {NoStop}%
\bibitem [{\citenamefont {Chern}\ \emph
  {et~al.}(2026{\natexlab{b}})\citenamefont {Chern}, \citenamefont {Fan},
  \citenamefont {Zhang},\ and\ \citenamefont {Zhang}}]{Chern2026b}%
  \BibitemOpen
  \bibfield  {author} {\bibinfo {author} {\bibfnamefont {G.-W.}\ \bibnamefont
  {Chern}}, \bibinfo {author} {\bibfnamefont {Y.}~\bibnamefont {Fan}}, \bibinfo
  {author} {\bibfnamefont {S.}~\bibnamefont {Zhang}},\ and\ \bibinfo {author}
  {\bibfnamefont {P.}~\bibnamefont {Zhang}},\ }\bibfield  {title} {\bibinfo
  {title} {Machine-learning modeling of magnetization dynamics in itinerant
  magnets},\ }\href {https://doi.org/10.1016/j.jmmm.2026.173679} {\bibfield
  {journal} {\bibinfo  {journal} {Journal of Magnetism and Magnetic Materials}\
  }\textbf {\bibinfo {volume} {628}},\ \bibinfo {pages} {173679} (\bibinfo
  {year} {2026}{\natexlab{b}})}\BibitemShut {NoStop}%
\bibitem [{\citenamefont {Mezard}\ and\ \citenamefont
  {Virasoro}(1987)}]{mezard87}%
  \BibitemOpen
  \bibfield  {author} {\bibinfo {author} {\bibfnamefont {P.~G.}\ \bibnamefont
  {Mezard}, \bibfnamefont {M.}}\ and\ \bibinfo {author} {\bibfnamefont
  {M.}~\bibnamefont {Virasoro}},\ }\href@noop {} {\emph {\bibinfo {title} {Spin
  Glass Theory and Beyond}}}\ (\bibinfo  {publisher} {World Scientific},\
  \bibinfo {address} {Singapore},\ \bibinfo {year} {1987})\BibitemShut
  {NoStop}%
\bibitem [{\citenamefont {Binder}\ and\ \citenamefont
  {Young}(1986)}]{binder86}%
  \BibitemOpen
  \bibfield  {author} {\bibinfo {author} {\bibfnamefont {K.}~\bibnamefont
  {Binder}}\ and\ \bibinfo {author} {\bibfnamefont {A.~P.}\ \bibnamefont
  {Young}},\ }\bibfield  {title} {\bibinfo {title} {Spin glasses: Experimental
  facts, theoretical concepts, and open questions},\ }\href
  {https://doi.org/10.1103/RevModPhys.58.801} {\bibfield  {journal} {\bibinfo
  {journal} {Rev. Mod. Phys.}\ }\textbf {\bibinfo {volume} {58}},\ \bibinfo
  {pages} {801} (\bibinfo {year} {1986})}\BibitemShut {NoStop}%
\bibitem [{\citenamefont {Mydosh}(1993)}]{mydosh95}%
  \BibitemOpen
  \bibfield  {author} {\bibinfo {author} {\bibfnamefont {J.~A.}\ \bibnamefont
  {Mydosh}},\ }\href@noop {} {\emph {\bibinfo {title} {Spin Glasses: An
  Experimental Introduction}}}\ (\bibinfo  {publisher} {Taylor \& Francis},\
  \bibinfo {address} {London},\ \bibinfo {year} {1993})\BibitemShut {NoStop}%
\bibitem [{\citenamefont {Mydosh}(2015)}]{mydosh15}%
  \BibitemOpen
  \bibfield  {author} {\bibinfo {author} {\bibfnamefont {J.~A.}\ \bibnamefont
  {Mydosh}},\ }\bibfield  {title} {\bibinfo {title} {Spin glasses: redux: an
  updated experimental/materials survey},\ }\href
  {https://doi.org/10.1088/0034-4885/78/5/052501} {\bibfield  {journal}
  {\bibinfo  {journal} {Reports on Progress in Physics}\ }\textbf {\bibinfo
  {volume} {78}},\ \bibinfo {pages} {052501} (\bibinfo {year}
  {2015})}\BibitemShut {NoStop}%
\bibitem [{\citenamefont {Cannella}\ and\ \citenamefont
  {Mydosh}(1972)}]{cannella72}%
  \BibitemOpen
  \bibfield  {author} {\bibinfo {author} {\bibfnamefont {V.}~\bibnamefont
  {Cannella}}\ and\ \bibinfo {author} {\bibfnamefont {J.~A.}\ \bibnamefont
  {Mydosh}},\ }\bibfield  {title} {\bibinfo {title} {Magnetic ordering in
  gold-iron alloys},\ }\href {https://doi.org/10.1103/PhysRevB.6.4220}
  {\bibfield  {journal} {\bibinfo  {journal} {Phys. Rev. B}\ }\textbf {\bibinfo
  {volume} {6}},\ \bibinfo {pages} {4220} (\bibinfo {year} {1972})}\BibitemShut
  {NoStop}%
\bibitem [{\citenamefont {Lamelas}\ \emph {et~al.}(1995)\citenamefont
  {Lamelas}, \citenamefont {Werner}, \citenamefont {Shapiro},\ and\
  \citenamefont {Mydosh}}]{lamelas95}%
  \BibitemOpen
  \bibfield  {author} {\bibinfo {author} {\bibfnamefont {F.~J.}\ \bibnamefont
  {Lamelas}}, \bibinfo {author} {\bibfnamefont {S.~A.}\ \bibnamefont {Werner}},
  \bibinfo {author} {\bibfnamefont {S.~M.}\ \bibnamefont {Shapiro}},\ and\
  \bibinfo {author} {\bibfnamefont {J.~A.}\ \bibnamefont {Mydosh}},\ }\bibfield
   {title} {\bibinfo {title} {Intrinsic spin-density-wave magnetism in cu-mn
  alloys},\ }\href {https://doi.org/10.1103/PhysRevB.51.621} {\bibfield
  {journal} {\bibinfo  {journal} {Phys. Rev. B}\ }\textbf {\bibinfo {volume}
  {51}},\ \bibinfo {pages} {621} (\bibinfo {year} {1995})}\BibitemShut
  {NoStop}%
\bibitem [{\citenamefont {Ruderman}\ and\ \citenamefont
  {Kittel}(1954)}]{ruderman54}%
  \BibitemOpen
  \bibfield  {author} {\bibinfo {author} {\bibfnamefont {M.~A.}\ \bibnamefont
  {Ruderman}}\ and\ \bibinfo {author} {\bibfnamefont {C.}~\bibnamefont
  {Kittel}},\ }\bibfield  {title} {\bibinfo {title} {Indirect exchange coupling
  of nuclear magnetic moments by conduction electrons},\ }\href
  {https://doi.org/10.1103/PhysRev.96.99} {\bibfield  {journal} {\bibinfo
  {journal} {Phys. Rev.}\ }\textbf {\bibinfo {volume} {96}},\ \bibinfo {pages}
  {99} (\bibinfo {year} {1954})}\BibitemShut {NoStop}%
\bibitem [{\citenamefont {Kasuya}(1956)}]{kasuya56}%
  \BibitemOpen
  \bibfield  {author} {\bibinfo {author} {\bibfnamefont {T.}~\bibnamefont
  {Kasuya}},\ }\bibfield  {title} {\bibinfo {title} {{A Theory of Metallic
  Ferro- and Antiferromagnetism on Zener's Model}},\ }\href
  {https://doi.org/10.1143/PTP.16.45} {\bibfield  {journal} {\bibinfo
  {journal} {Progress of Theoretical Physics}\ }\textbf {\bibinfo {volume}
  {16}},\ \bibinfo {pages} {45} (\bibinfo {year} {1956})}\BibitemShut {NoStop}%
\bibitem [{\citenamefont {Yosida}(1957)}]{yosida57}%
  \BibitemOpen
  \bibfield  {author} {\bibinfo {author} {\bibfnamefont {K.}~\bibnamefont
  {Yosida}},\ }\bibfield  {title} {\bibinfo {title} {Magnetic properties of
  {C}u-{M}n alloys},\ }\href {https://doi.org/10.1103/PhysRev.106.893}
  {\bibfield  {journal} {\bibinfo  {journal} {Phys. Rev.}\ }\textbf {\bibinfo
  {volume} {106}},\ \bibinfo {pages} {893} (\bibinfo {year}
  {1957})}\BibitemShut {NoStop}%
\bibitem [{\citenamefont {Zener}(1951)}]{Zener1951a}%
  \BibitemOpen
  \bibfield  {author} {\bibinfo {author} {\bibfnamefont {C.}~\bibnamefont
  {Zener}},\ }\bibfield  {title} {\bibinfo {title} {Interaction between the d
  shells in the transition metals},\ }\href
  {https://doi.org/10.1103/PhysRev.81.440} {\bibfield  {journal} {\bibinfo
  {journal} {Physical Review}\ }\textbf {\bibinfo {volume} {81}},\ \bibinfo
  {pages} {440} (\bibinfo {year} {1951})}\BibitemShut {NoStop}%
\bibitem [{\citenamefont {de~Gennes}(1960)}]{deGennes1960}%
  \BibitemOpen
  \bibfield  {author} {\bibinfo {author} {\bibfnamefont {P.-G.}\ \bibnamefont
  {de~Gennes}},\ }\bibfield  {title} {\bibinfo {title} {Effects of double
  exchange in magnetic crystals},\ }\href
  {https://doi.org/10.1103/PhysRev.118.141} {\bibfield  {journal} {\bibinfo
  {journal} {Physical Review}\ }\textbf {\bibinfo {volume} {118}},\ \bibinfo
  {pages} {141} (\bibinfo {year} {1960})}\BibitemShut {NoStop}%
\bibitem [{\citenamefont {Anderson}(1961)}]{anderson61}%
  \BibitemOpen
  \bibfield  {author} {\bibinfo {author} {\bibfnamefont {P.~W.}\ \bibnamefont
  {Anderson}},\ }\bibfield  {title} {\bibinfo {title} {Localized magnetic
  states in metals},\ }\href {https://doi.org/10.1103/PhysRev.124.41}
  {\bibfield  {journal} {\bibinfo  {journal} {Phys. Rev.}\ }\textbf {\bibinfo
  {volume} {124}},\ \bibinfo {pages} {41} (\bibinfo {year} {1961})}\BibitemShut
  {NoStop}%
\bibitem [{\citenamefont {He}\ \emph {et~al.}(2016)\citenamefont {He},
  \citenamefont {Zhang}, \citenamefont {Ren},\ and\ \citenamefont
  {Sun}}]{He2016}%
  \BibitemOpen
  \bibfield  {author} {\bibinfo {author} {\bibfnamefont {K.}~\bibnamefont
  {He}}, \bibinfo {author} {\bibfnamefont {X.}~\bibnamefont {Zhang}}, \bibinfo
  {author} {\bibfnamefont {S.}~\bibnamefont {Ren}},\ and\ \bibinfo {author}
  {\bibfnamefont {J.}~\bibnamefont {Sun}},\ }\bibfield  {title} {\bibinfo
  {title} {Deep residual learning for image recognition},\ }in\ \href
  {https://doi.org/10.1109/CVPR.2016.90} {\emph {\bibinfo {booktitle}
  {Proceedings of the IEEE Conference on Computer Vision and Pattern
  Recognition (CVPR)}}}\ (\bibinfo {year} {2016})\ pp.\ \bibinfo {pages}
  {770--778}\BibitemShut {NoStop}%
\bibitem [{\citenamefont {Landau}\ and\ \citenamefont
  {Lifshitz}(1935)}]{Landau1935}%
  \BibitemOpen
  \bibfield  {author} {\bibinfo {author} {\bibfnamefont {L.}~\bibnamefont
  {Landau}}\ and\ \bibinfo {author} {\bibfnamefont {E.}~\bibnamefont
  {Lifshitz}},\ }\bibfield  {title} {\bibinfo {title} {On the theory of the
  dispersion of magnetic permeability in ferromagnetic bodies},\ }\href@noop {}
  {\bibfield  {journal} {\bibinfo  {journal} {Physikalische Zeitschrift der
  Sowjetunion}\ }\textbf {\bibinfo {volume} {8}},\ \bibinfo {pages} {153}
  (\bibinfo {year} {1935})}\BibitemShut {NoStop}%
\bibitem [{\citenamefont {Gilbert}(2004)}]{Gilbert2004}%
  \BibitemOpen
  \bibfield  {author} {\bibinfo {author} {\bibfnamefont {T.~L.}\ \bibnamefont
  {Gilbert}},\ }\bibfield  {title} {\bibinfo {title} {A phenomenological theory
  of damping in ferromagnetic materials},\ }\href
  {https://doi.org/10.1109/TMAG.2004.836740} {\bibfield  {journal} {\bibinfo
  {journal} {IEEE Transactions on Magnetics}\ }\textbf {\bibinfo {volume}
  {40}},\ \bibinfo {pages} {3443} (\bibinfo {year} {2004})}\BibitemShut
  {NoStop}%
\bibitem [{\citenamefont {Evans}\ \emph {et~al.}(2014)\citenamefont {Evans},
  \citenamefont {Fan}, \citenamefont {Chureemart}, \citenamefont {Ostler},
  \citenamefont {Ellis},\ and\ \citenamefont {Chantrell}}]{Evans2014}%
  \BibitemOpen
  \bibfield  {author} {\bibinfo {author} {\bibfnamefont {R.~F.~L.}\
  \bibnamefont {Evans}}, \bibinfo {author} {\bibfnamefont {W.~J.}\ \bibnamefont
  {Fan}}, \bibinfo {author} {\bibfnamefont {P.}~\bibnamefont {Chureemart}},
  \bibinfo {author} {\bibfnamefont {T.~A.}\ \bibnamefont {Ostler}}, \bibinfo
  {author} {\bibfnamefont {M.~O.~A.}\ \bibnamefont {Ellis}},\ and\ \bibinfo
  {author} {\bibfnamefont {R.~W.}\ \bibnamefont {Chantrell}},\ }\bibfield
  {title} {\bibinfo {title} {Atomistic spin model simulations of magnetic
  nanomaterials},\ }\href {https://doi.org/10.1088/0953-8984/26/10/103202}
  {\bibfield  {journal} {\bibinfo  {journal} {Journal of Physics: Condensed
  Matter}\ }\textbf {\bibinfo {volume} {26}},\ \bibinfo {pages} {103202}
  (\bibinfo {year} {2014})}\BibitemShut {NoStop}%
\bibitem [{\citenamefont {Feynman}(1939)}]{Feynman1939}%
  \BibitemOpen
  \bibfield  {author} {\bibinfo {author} {\bibfnamefont {R.~P.}\ \bibnamefont
  {Feynman}},\ }\bibfield  {title} {\bibinfo {title} {Forces in molecules},\
  }\href {https://doi.org/10.1103/PhysRev.56.340} {\bibfield  {journal}
  {\bibinfo  {journal} {Physical Review}\ }\textbf {\bibinfo {volume} {56}},\
  \bibinfo {pages} {340} (\bibinfo {year} {1939})}\BibitemShut {NoStop}%
\bibitem [{\citenamefont {Kingma}\ and\ \citenamefont {Ba}(2017)}]{kingma17}%
  \BibitemOpen
  \bibfield  {author} {\bibinfo {author} {\bibfnamefont {D.~P.}\ \bibnamefont
  {Kingma}}\ and\ \bibinfo {author} {\bibfnamefont {J.}~\bibnamefont {Ba}},\
  }\href@noop {} {\bibinfo {title} {Adam: A method for stochastic
  optimization}} (\bibinfo {year} {2017}),\ \Eprint
  {https://arxiv.org/abs/1412.6980} {arXiv:1412.6980 [cs.LG]} \BibitemShut
  {NoStop}%
\end{thebibliography}%

\end{document}